\documentclass{aastex631}
\usepackage{graphicx}
\usepackage{amsmath}
\usepackage{natbib}
\usepackage{amssymb}
\usepackage{sidecap}
\usepackage{booktabs}
\usepackage[T1]{fontenc} 

\newcommand\yvette{C24}
\newcommand\vlassnum{91}

\shorttitle{Delayed Radio TDEs II}
\shortauthors{Alexander et al.}
\graphicspath{{./}{figures/}}

\begin{document}

\title{The Multi-Wavelength Context of Delayed Radio Emission in TDEs: Evidence for Accretion-Driven Outflows}

\author[0000-0002-8297-2473]{Kate D. Alexander}
\affiliation{Department of Astronomy and Steward Observatory, University of Arizona, 933 North Cherry Avenue, Tucson, AZ 85721-0065, USA}

\author[0000-0003-4768-7586]{Raffaella Margutti}
\affiliation{Department of Astronomy, University of California, Berkeley, CA 94720-3411, USA}

\author[0000-0001-6395-6702]{Sebastian Gomez}
\affiliation{Center for Astrophysics | Harvard \& Smithsonian, Cambridge, MA 02138, USA}

\author[0000-0002-3019-4577]{Michael Stroh}
\affiliation{Center for Interdisciplinary Exploration and Research in Astrophysics (CIERA) and Department of Physics and Astronomy, Northwestern University, 1800 Sherman Ave,Evanston, IL 60201,USA}

\author[0000-0002-7706-5668]{Ryan~Chornock}
\affiliation{Department of Astronomy, University of California, Berkeley, CA 94720-3411, USA}

\author[0000-0003-1792-2338]{Tanmoy Laskar}
\affiliation{Department of Physics \& Astronomy, University of Utah, Salt Lake City, UT 84112, USA}
\affiliation{Department of Astrophysics/IMAPP, Radboud University, PO Box 9010, 6500 GL Nijmegen, The Netherlands}

\author[0000-0001-7007-6295]{Y. Cendes}
\affiliation{Department of Physics, University of Oregon, 1371 E 13th Ave, Eugene OR 97403, USA}
\affiliation{Institute for Fundamental Science, University of Oregon, 1371 E 13th Ave, Eugene OR 97403, USA}

\author[0000-0002-9392-9681]{Edo Berger}
\affiliation{Center for Astrophysics | Harvard \& Smithsonian, Cambridge, MA 02138, USA}

\author[0000-0003-0307-9984]{Tarraneh Eftekhari}
\affiliation{Center for Interdisciplinary Exploration and Research in Astrophysics (CIERA), Northwestern University, 1800 Sherman Ave, Evanston, IL 60201,USA}

\author[0000-0003-4537-3575]{Noah Franz}
\affiliation{Department of Astronomy and Steward Observatory, University of Arizona, 933 North Cherry Avenue, Tucson, AZ 85721-0065, USA}

\author[0000-0003-2349-101X]{Aprajita Hajela}
\affiliation{DARK, Niels Bohr Institute, University of Copenhagen, Jagtvej 155, 2200 Copenhagen, Denmark}

\author[0000-0002-4670-7509]{B. D. Metzger}
\affil{Department of Physics and Columbia Astrophysics Laboratory, Columbia University, Pupin Hall, New York, NY 10027, USA}
\affil{Center for Computational Astrophysics, Flatiron Institute, 162 5th Ave, New York, NY 10010, USA} 

\author[0000-0003-0794-5982]{Giacomo Terreran}
\affiliation{Adler Planetarium, 1300 S Lake Shore Dr, Chicago, IL 60605, USA}

\author[0000-0002-0592-4152]{Michael Bietenholz}
\affil{SARAO/Hartebeesthoek Radio Astronomy Observatory, PO Box 443, Krugersdorp, 1740, South Africa}
\affil{Department of Physics and Astronomy, York University, Toronto, M3J 1P3, Ontario, Canada}

\author[0000-0003-0528-202X]{Collin Christy}
\affiliation{Department of Astronomy and Steward Observatory, University of Arizona, 933 North Cherry Avenue, Tucson, AZ 85721-0065, USA}

\author[0000-0002-3137-4633]{Fabio De Colle}
\affiliation{Instituto de Ciencias Nucleares, Universidad Nacional Aut\'onoma de M\'exico, A. P. 70-543 04510 D. F. Mexico}

\author[0000-0003-4183-4215]{S.~Komossa}
\affiliation{Max-Planck-Institut f\"ur Radiostronomie, Auf dem H\"ugel 69, 53121 Bonn,
Germany}

\author[0000-0002-2555-3192]{Matt Nicholl}
\affiliation{Astrophysics Research Centre, School of Mathematics and Physics, Queens University Belfast, Belfast BT7 1NN, UK}

\author[0000-0003-2558-3102]{Enrico Ramirez-Ruiz}
\affiliation{Department of Astronomy and Astrophysics, University of California, Santa Cruz, CA 95064, USA}

\author[0000-0002-4912-2477]{Richard Saxton}
\affiliation{Telespazio UK for ESA, ESAC, Apartado 78, 28691 Villanueva de la Cañada, Madrid, Spain}

\author[0000-0001-9915-8147]{Genevieve Schroeder}
\affiliation{Department of Astronomy, Cornell University, Ithaca, NY 14853, USA}

\author[0000-0003-3734-3587]{Peter K. G. Williams}
\affiliation{Center for Astrophysics | Harvard \& Smithsonian, Cambridge, MA 02138, USA}

\author[0009-0000-0753-4345]{William Wu}
\affiliation{Department of Astronomy, University of California, Berkeley, CA 94720-3411, USA}



\begin{abstract}

Recent observations presented in \cite{cba+23} show that optically selected tidal disruption events (TDEs) commonly produce delayed radio emission that can peak years post-disruption. Here, we explore the multi-wavelength properties of a sample of radio-observed optically selected TDEs to shed light on the physical process(es) responsible for the late-rising radio emission. We combine new late-time X-ray observations with archival optical, UV, X-ray, and radio data to conclude that a diversity of accretion-driven outflows may power delayed radio emission in TDEs.
Our analysis suggests that some late radio outflows may be launched by a delayed phase of super-Eddington accretion onto the central supermassive black hole (SMBH), while others may result from a state transition to a ``low-hard'' radiatively inefficient accretion flow or the deceleration of an off-axis relativistic jet. {We find that TDEs with delayed radio emission are less likely to exhibit helium emission lines at early times ($p=0.002$) and may have larger optical/UV photospheric radii ($p=0.026$) than other TDEs,} possibly also indicating that the onset of SMBH accretion is delayed in these systems. Our results have implications for our understanding of state changes in SMBH accretion flows, the circularization timescale for TDE debris, and the prevalence of off-axis jets in TDEs, and motivate systematic, long-term monitoring of these unique transients. The objects in our sample {with the brightest radio emission} are also detected in the VLA Sky Survey (VLASS), demonstrating that all-sky radio surveys can play an important role in discovering unexpected properties of the TDE population.

\end{abstract}



\section{Introduction} \label{sec:intro}

The strong gravitational field near a supermassive black hole (SMBH) is capable of tidally disrupting an approaching star, temporarily enhancing the mass fallback rate onto the SMBH by orders of magnitude as gas from the star is accreted \citep{h75,r88}. Non-thermal synchrotron emission from these tidal disruption events (TDEs) is best observed in the radio or millimeter and has been variously ascribed to jets, non-relativistic outflows, interactions between unbound stellar debris and the circumnuclear medium, or some combination thereof (see \citealt{avhz20} for a review). A small fraction ($\lesssim1$\%) of known TDEs produce powerful relativistic jets with peak radio luminosities $\nu L_{\nu}\gtrsim10^{41}$ erg s$^{-1}$ \citep{zbs+11,ckh+12,bls+15,acp+22}, but most TDEs lack such jets and are instead discovered as long-lasting transients in X-ray or optical surveys (e.g., \citealt{bkd96,skaj20,sgm+21, vgh+21,hvg+23,yrg+23}). Targeted radio follow up of individual X-ray and optically discovered TDEs has revealed that some such TDEs exhibit lower-luminosity radio emission ($\nu L_{\nu}\lesssim10^{39}$ erg s$^{-1}$) within a few months of discovery (e.g., \citealt{vas+16,abg+16, awb+17,svk+21,cab+21}), but early follow up efforts typically ceased if no radio emission was detected within $6-12$ months, leaving the late-time radio properties of most TDEs unknown.

The first searches for TDE radio emission on longer timescales of months to years post-discovery were optimized to target initially-relativistic jets viewed off-axis, as the radio emission of such jets would be suppressed at early times due to relativistic beaming and only become visible after jet deceleration. These searches resulted in only non-detections, confirming that the rarity of powerful relativistic jets was a physical and not a selection effect; however, these studies were not sensitive enough to probe the existence of weaker outflows on similar timescales \citep{k02,vfkf13,bmc+13}. Recently, deeper wide-field radio transient searches have resulted in the first blindly selected radio TDE candidates (i.e., nuclear radio transients discovered independently, without prior detection of a counterpart at other wavelengths), suggesting that fainter radio emission in TDEs may be $\sim10$ times more common than powerful jets (\citealt{amh+20}; see also \citealt{srd+23b,ddg+24}). However, it is still not clear if these radio-selected nuclear transients probe the same underlying population as optically selected or X-ray selected TDEs. 

This is the second of two papers comprising the first large-scale attempt to quantify the prevalence of {less luminous}, late-rising radio emission in TDEs and understand the physical origin of this emission. The first of the two papers, \citet{cba+23} (hereafter \yvette), presented the surprising result that \emph{late-rising radio emission is common in optically selected TDEs,} with $\sim40$\% of the observed objects showing variable radio emission years post-discovery. This is comparable to --- or possibly slightly higher than --- the rate of TDEs that exhibit ``prompt'' radio emission (i.e., within 6 months of discovery), although the sample size of TDEs with early radio observations remains small and the occurrence rate of prompt radio emission in TDEs thus remains poorly constrained \citep{avhz20}. The delayed radio flares characterized by \yvette\ and others (e.g., \citealt{hca21, hsf+21, cba+22, ham+24}) are generally faint ($\nu L_{\nu}\lesssim10^{39}$ erg s$^{-1}$, with a few exceptions) and have been variously ascribed to non-relativistic outflows launched several hundred days post-disruption \citep{hca21,cba+22}, to off-axis relativistic jets launched promptly \citep{mp23,sbh+24}, to initially-choked precessing relativistic jets that only escape after an initially misaligned accretion disk aligns with the black hole spin axis \citep{tm23,lmm24}, or to outflowing material encountering changes in the density of the ambient circumnuclear medium \citep{mp24,zsml24}. While the physical origin of TDE radio emission remains unclear, the wide range of emission onset times  --- from a few days post-discovery to $>1000$ days post-discovery --- may suggest that multiple mechanisms are responsible for producing low-luminosity radio emission in TDEs. 

The multi-wavelength properties of TDEs may shed light on the origin of late-rising radio emission, as they give us independent methods to trace the evolution of the SMBH accretion rate with time. Mapping the accretion rate has been attempted in a variety of ways, most commonly by modeling TDEs' optical/UV or X-ray light curves (e.g., \citealt{mgr19,skaj20}). If delayed radio brightenings are caused by newly-launched accretion-driven outflows, then we expect some correlation between the time when radio emission first appears and the instantaneous accretion rate onto the SMBH at that time. Conversely, if the delayed radio brightenings are powered by a promptly-launched outflow encountering changes in the ambient environment, we would not expect such correlations. 

While the optical through X-ray emission from TDEs is generally broadly consistent with thermal emission from hot gas near the SMBH, the exact connection between the observed optical, UV, and X-ray light curves of TDEs and the accretion rate has been the subject of intense theoretical debate {(e.g., \citealt{psk+15,skc+15,rkgr16,dmr+18,mgr19,lb20,rrk+20} and references therein; \citealt{rkpn20,m22,sm24})}. One open question is whether the mass fallback rate $\dot{M}_{\rm fb}$ (i.e., the rate at which stellar debris approaches the black hole from the point of disruption) maps directly to the accretion rate $\dot{M}$ (i.e., the rate at which material crosses the event horizon), or if the latter is significantly delayed relative to the former. {The mass fallback rate may be easier to constrain observationally in some cases, as it is generally assumed to be directly proportional to the bolometric luminosity, $L=\epsilon\dot{M}_{\rm fb}c^2$, where $\epsilon$ is a constant (typically $\sim0.1$). However, it is the accretion rate that is relevant for predicting the launch times of radio-producing outflows. Measuring the accretion rate is less straightforward due to the inherent messiness of the disruption process and debris circularization.} For this work, we therefore consider two extremal possibilities; namely Case 1: that the stellar debris rapidly circularizes and forms an accretion flow, such that the bolometric optical-through-UV light curve directly traces {both} the mass fallback rate (typically expected to exceed the Eddington limit before declining as $t^{-5/3}$; \citealt{r88}) {and} the instantaneous accretion rate onto the SMBH (accounting for a minimal viscous delay, e.g., \citealt{gr13,mgr19}); or Case 2: that the peak accretion rate onto the SMBH is significantly delayed relative to peak optical light, either because the debris circularization is inefficient (in which case the optical peak may be powered instead by shocks from stream-stream collisions, e.g., \citealt{psk+15,skc+15,lb20}) or because radiation pressure initially prevents even rapidly circularized debris from collapsing into a disk while the mass fallback rate is high, instead forming a hot, extended quasi-spherical pressure-supported envelope that powers the early optical emission while keeping the accretion rate across the event horizon low \citep{lu97,gmr14,m22}. In Case 1, X-rays produced near the SMBH are reprocessed into the optical band by more extended material, so the accretion rate is best traced in the optical/UV {and can be computed by fitting the optical/UV light curves with existing modeling codes} \citep{cb14,gmr14,ms16,rkgr16,dmr+18,mgr19}. In Case 2, tracking the X-ray luminosity {and comparing it to model predictions} may provide a better method of estimating the instantaneous accretion rate, as thermal X-ray emission in these models originates in the innermost part of the accretion flow and thus indicates that matter is successfully approaching the SMBH regardless of the origin of the optical emission (\citealt{agr17,gca17,m22}; {\citealt{sm24}}). In both cases, state transitions in the accretion flow are expected as the accretion rate changes, causing the production and quenching of outflows, but the timing of these transitions relative to peak optical light will be different. We compare the onset of any observed radio emission to the expected timing of state transitions in the accretion flow in both cases, focusing on the highly super-Eddington and highly sub-Eddington accretion phases that are most likely to power outflows (e.g., \citealt{gm11,dgnr12}).

\yvette\ modeled the radio emission detected in optically selected TDEs at late times using an equipartition analysis, discussed its properties in the context of other radio observations of TDEs, and calculated the occurrence rate of late-time radio emission. Here, we discuss the multi-wavelength properties of radio-observed optical TDEs in more detail and attempt to discriminate among possible physical models for delayed radio emission. In Section \ref{sec:sample} we introduce our sample and describe our new multi-wavelength observations and the archival data that we use for our subsequent analysis. We describe our light curve modeling and compare the radio properties of our sample to their multi-wavelength emission and derived properties in Section \ref{sec:mod}. We discuss the implications of our results in Section \ref{sec:disc} and we conclude in Section \ref{sec:conc}.

\section{Sample definition and data collection}\label{sec:sample}

This work was originally motivated by a search for radio emission associated with \vlassnum\ known TDE candidates in the VLA Sky Survey (VLASS). This search resulted in three detections of transient radio emission associated with high-confidence TDEs and three detections of likely unrelated radio sources (Appendix \ref{sec:radio}). Surprisingly, the detected TDEs included a dramatic rebrightening of the optical TDE ASASSN-15oi discovered in VLASS epoch 1 at $\sim3$ years post-disruption \citep{hca21,ham+24}, motivating the deeper radio searches presented in \yvette\ and this paper. A second optical TDE in our sample, AT2018hyz, appears as an even brighter radio transient in VLASS epoch 3 (\citealt{cba+22}; \yvette).

To probe the physical mechanism or mechanisms behind the newly discovered late-rising radio emission in optically detected TDEs, we utilize multi-wavelength observations of a sample of TDEs discovered in optical surveys between 2014 and 2020 (Table \ref{tab:gold}). Our sample contains 31 TDEs for which targeted radio follow up was obtained. {Most are at redshift $z\lesssim0.1$; we also include two TDEs at slightly higher $z$ with constraining early radio data (DES14C1kia and AT2020opy).} As we wish to understand if late-time radio flares have a different physical origin than prompt radio emission in TDEs, our sample consists of the 24 TDEs observed by \yvette, most of which had no published radio detections prior to that work, plus an additional 7 TDEs from the literature previously reported to exhibit radio emission at early times. In addition to compiling optical, UV, and X-ray photometry of our TDEs from the literature, we also obtained new late-time X-ray observations of all TDEs in our sample. We briefly summarize all data used in this paper below and provide details of our new data reduction and analysis. A complete list of all references for previously-published data used in this work is given in Appendix \ref{sec:data} (Table \ref{tab:data}), along with all new radio and X-ray photometry used for our analysis (Tables \ref{tab:19qiz}, \ref{tab:xrays}). The complete multiwavelength dataset will also be made available as part of the Open mulTiwavelength Transient Event Repository (OTTER; {\citealt{fag+25}}).

\begin{deluxetable*}{llllcll}
\tablenum{1}
\tablecaption{Our Sample: Optically-selected TDEs at $z\lesssim0.1$ with radio observations discovered between January 2014 and October 2020. TDEs in bold are those with newly discovered radio emission, presented in full in \yvette. 
\label{tab:gold}}
\tablewidth{0pt}
\tablehead{
\colhead{TDE Name} & \colhead{Redshift} & \colhead{Discovery} & \colhead{Date of first} & \colhead{$\Delta t$ of first radio} & \colhead{Radio} & \colhead{Optical} \\ 
\nocolhead{Name} & \nocolhead{Number} & \colhead{Date} & \colhead{radio detection} &\colhead{detection (rest-frame days)} & \colhead{Type} & \colhead{Type}  
}
\startdata
\textbf{ASASSN-14ae}$^\dagger$ & 0.0436 & 2014 Jan 25 & 2020 May 25 & 2216 & Delayed & {TDE-H}$^\dagger$ \\ 
DES14C1kia & 0.162 & 2014 Nov 11 & None & -- & None & TDE-He \\ 
ASASSN-14li & 0.02058 & 2014 Nov 22 & 2014 Dec 24 & 44 & Prompt & TDE-H+He \\ 
iPTF15af & 0.07897 & 2015 Jan 15 & None & -- & None & TDE-H+He \\ 
ASASSN-15oi & 0.0484 & 2015 Aug 14 & 2016 Feb 12 & 177 & Delayed & TDE-He \\ 
iPTF16axa & 0.108 & 2016 May 29 & None & -- & None & TDE-H+He \\ 
\textbf{PS16dtm (AT2016ezh)} & 0.0804 & 2016 Aug 12 & 2020 Jun 6  & 1287 & Delayed & TDE-H \\ 
iPTF16fnl (AT2016fnl)$^\ddagger$ & 0.016328 & 2016 Aug 29 & 2017 Jan 29 & 151 & Prompt$^\ddagger$ & TDE-H+He \\ 
\textbf{OGLE17aaj*} & 0.116 & 2017 Jan 2 & 2021 May 1 & 1417* & Host* & TDE-H+He \\ 
AT2017eqx & 0.1089 & 2017 Jun 7 & None & -- & None & TDE-H+He \\ 
\textbf{AT2018zr}$^\dagger$  & 0.071 & 2018 Mar 2 & 2022 Oct 16 & 1599 & Delayed & {TDE-H}$^\dagger$ \\ 
\textbf{AT2018bsi*} & 0.051 & 2018 Apr 9 & 2022 Dec 8 & 1622* & Host* & TDE-H+He \\ 
\textbf{AT2018dyb} & 0.018 & 2018 Jul 11 & 2021 May 4 & 1010 & Delayed & TDE-H+He \\ 
AT2018fyk & 0.059 & 2018 Sep 8 & 2024 Feb 20 & 1880 & Delayed & TDE-H+He \\ 
\textbf{AT2018hco} & 0.088 & 2018 Oct 4 & 2021 Jun 12 & 903 & Delayed & TDE-H \\ 
AT2018hyz$^\dagger$ & 0.04573 & 2018 Nov 6 & 2021 Jun 11 & 934 & Delayed & {TDE-H}$^\dagger$ \\ 
AT2018lna & 0.091 & 2018 Dec 28 & None & -- & None & TDE-H+He \\ 
ASASSN-19bt (AT2019ahk) & 0.0262 & 2019 Jan 29 & 2019 Mar 3 & 39 & Prompt & TDE-H \\ 
AT2019azh & 0.0222 & 2019 Feb 22 & 2019 Mar 15 & 21 & Prompt & TDE-H+He \\ 
AT2019dsg & 0.0512 & 2019 Apr 9 & 2019 May 22  & 43 & Prompt & TDE-H+He \\ 
\textbf{AT2019ehz} & 0.074 & 2019 Apr 29 & 2021 Jun 11  & 722 & Delayed & TDE-H \\ 
\textbf{AT2019eve} & 0.064 & 2019 May 5 & 2021 Jun 11  & 723 & Delayed & TDE-H \\ 
AT2019qiz & 0.0151 & 2019 Sep 19 & 2019 Sep 26  & 6 & Prompt & TDE-H+He \\ 
\textbf{AT2019teq} & 0.0878 & 2019 Oct 20 & 2022 Oct 19 & 1008 & Delayed & TDE-H+He \\ 
\textbf{AT2020pj*} & 0.068 & 2020 Jan 2 & 2022 Oct 28 & 964* & Host* & TDE-H+He \\ 
AT2020mot* & 0.07 & 2020 Jun 14 & 2021 Jan 29 & 214* & Host* & TDE-H+He \\ 
\textbf{AT2020neh} & 0.062 & 2020 Jun 19 & 2022 Nov 8 & 823 & Delayed & TDE-H+He \\ 
\textbf{AT2020nov*} & 0.084 & 2020 Jun 27 & 2020 Oct 16 & 102* & Host* & TDE-H+He \\ 
AT2020opy & 0.159 & 2020 Jul 8 & 2020 Oct 6 & 85 & Prompt & TDE-H+He \\ 
\textbf{AT2020wey*} & 0.02738 & 2020 Oct 8 & 2021 Feb 13 & 125* & Host* & TDE-H+He \\ 
AT2020vwl & 0.035 & 2020 Oct 10 & 2021 Feb 23  & 118 & Prompt & TDE-H+He \\ 
\enddata
\smallskip
*indicates origin of radio emission is either ambiguous or unrelated to the TDE.

$^\dagger${\yvette\ classify these TDEs as TDE-H+He because they exhibit He emission at $\gtrsim50-200$ days. However, we follow other authors (e.g., \citealt{vgh+21,hcg+23}) in classifying them as TDE-H because they showed only H lines in their initial classification spectra around peak optical light.}

$^\ddagger$iPTF16fnl has been previously referred to in the literature as a ``delayed'' radio TDE due to several radio upper limits at early times \citep{hsf+21}. However, its first radio detection is $153$ $(151)$ days post-discovery in the observer (rest) frame, which places it in the ``prompt'' category as defined here (Section \ref{sec:rdata}).
\end{deluxetable*}

\subsection{Radio: VLA, MeerKAT, ATCA, and literature data}\label{sec:rdata}

\yvette\ presented new radio observations at 2 -- 6 years post-discovery of 24 optically-selected TDEs at $z \lesssim 0.1$ discovered between January 2014 and October 2020\@. These 24 TDEs are drawn from the samples defined by \cite{vgh+21} and \cite{hvg+23}, plus a few additional events from the literature \citep{fbc+15,Blanchard17,ghw+19,fgv+20,abm+22}. Following \yvette, we also include previously-published radio observations and public archival data from the VLASS and VAST radio surveys \citep{vlass,vast} when available. For this work, we reanalyzed one observation of PS16dtm previously reported in \yvette, taken with NSF's Karl G. Jansky Very Large Array (the VLA) on 2017 Aug 22 (372 days post-discovery) under program 16B-318 (PI: Alexander), and report a tentative new detection of $39\pm10$ $\mu$Jy at 6 GHz. For AT2018fyk, we also include the recent radio detections reported by \cite{cba+24b}. 

In this paper, we add 7 TDEs from the literature with promptly-detected radio emission, to compare and contrast with the late-rising radio TDEs. Six of these (ASASSN-14li, ASASSN-15oi, AT2019azh, ASASSN-19bt, AT2019qiz, and AT2020opy) are included in the population analyses by \cite{vgh+21} and/or \cite{hvg+23}; the remaining TDE is the youngest object in our sample and the first TDE classified as part of the second phase of the Zwicky Transient Facility survey (ZTF-II; \citealt{ztf}), AT2020vwl \citep{yrg+23,gam+23}. 
For AT2019qiz, we also include a subset of the VLA observations collected under programs 19A-013, 20A-372 (PI: Alexander), and 21A-303 (PI: Hajela), a public observation collected as part of the VLASS epoch 3, and a MeerKAT observation collected under program SCI-20220822-YC-01 (PI: Cendes). The AT2019qiz radio data used in our analysis are listed in Table \ref{tab:19qiz} in Appendix \ref{sec:data}, and the complete radio dataset for AT2019qiz will be presented in future work (Franz et al.\ in prep).

All VLA data were reduced in CASA \citep{casa} using standard procedures. For the MeerKAT observation, we used the standard calibrated MeerKAT pipeline images available via the SARAO Science Data Processor (SDP)\footnote{https://skaafrica.atlassian.net/wiki/spaces/ESDKB/pages/338723406/}. All flux densities were measured using the {\tt imtool} package within {\tt pwkit} \citep{pwkit}, which fits a Gaussian within a small box centered on the pixel coordinates provided by the user. This Gaussian can either be fixed to be the same size and shape as the telescope beam for a point source fit, or allowed to vary freely for the case of extended emission. (By default {\tt imtool} will compute the flux density both ways and return the model that provides a better fit/lower residuals, using a similar algorithm to the MIRIAD task {\tt imfit}.) For virtually all of the sources in this paper, we do not expect significant contamination from extended host galaxy emission due to the relatively large distances and low star formation rates of the TDE hosts. We therefore use a point source fit, as indeed preferred for our images. 

The full list of literature sources used to compile our radio dataset for each TDE in our sample is given in Appendix \ref{sec:data} (Table \ref{tab:data}). For the purpose of our analysis, we classify each TDE's radio emission as ``Prompt:'' first radio detection $<6$ months post-discovery (8 TDEs); ``Delayed:'' first radio detection $>6$ months post-discovery (12 TDEs), ``None:'' all radio observations to date resulted in non-detections (5 TDEs), or ``Host:'' persistent radio emission likely unrelated to the TDE (6 TDEs). The ``Host'' classification also includes TDEs where the origin of the observed radio emission is ambiguous (e.g., because the emission is too faint to definitively assess variability). 
We classify TDEs that exhibit multiple radio peaks (ASASSN-15oi and AT2019dsg) based on the onset time of the brightest radio emission episode. {We note that some TDEs in our sample are significantly less well-observed in the radio than others. In particular,} several of the TDEs that we classify as radio type Delayed have no radio observations at $t<6$ months, meaning that we have no way of knowing if they also displayed prompt radio emission and are thus misclassified here, potentially diluting any underlying trends (Section \ref{sec:super}). The radio classification of each TDE is given in Table \ref{tab:gold}.

\subsection{X-rays: archival data and new Swift-XRT observations}\label{sec:xrays}

All of the TDEs in our sample have received at least some X-ray follow-up, although the cadence and data quality vary significantly. We provide full details of the literature X-ray observations used in Table \ref{tab:data} in Appendix \ref{sec:data}. To systematically explore the late-time X-ray properties of our sample, we triggered new target-of-opportunity (TOO) observations with the X-ray Telescope (XRT, \citealt{Burrows05}) onboard the Neil Gehrels Swift Observatory \citep{Gehrels04} of all TDEs in Table \ref{tab:gold} without existing late-time X-ray coverage, which cover the timescale $\sim2-8$ yrs since discovery. We also re-analyzed all the available Swift-XRT data of TDEs in our sample not already reported in the literature, using the online automated tools released by the Swift-XRT team \citep{Evans09}\footnote{\texttt{https://www.swift.ac.uk/user\_objects/}} and custom scripts \citep{Margutti13}. We first extracted a 0.3--10 keV count-rate light-curve 
with a minimum number of counts per bin allowing for dynamic binning.  In many cases, these are TDEs with only faint X-rays or non-detections, so we assume the same spectral model for all TDEs for the flux calibration and {fix $N_{\rm H,int}=0$ (meaning that we assume zero $N_{\rm H}$ in the TDE host galaxy and we account only for the Galactic $N_{\rm H}$ expected along the line of sight)}. We considered two spectral models: a constant non-thermal power-law spectrum with photon index $\Gamma=2$; and a constant thermal spectrum consisting of a black-body model with $k_{b}T=0.05\,\rm{keV}$.   These spectral models and parameters are broadly representative of the hard (power-law) and soft (black-body) X-ray spectra of TDEs \citep{skaj20,ggy+23}. For the power-law
and black-body models, the average count-to-flux factors are $\approx 5\times 10^{-11}\rm{erg\,s^{-1}cm^{-2}}/\rm{c\,s^{-1}}$ and $\approx 3.3\times 10^{-11}\rm{erg\,s^{-1}cm^{-2}}/\rm{c\,s^{-1}}$, respectively.  In the following we employ the power-law model flux calibration for all the TDE X-ray light-curves; we emphasize that the difference of a factor $<2$ between the two flux calibrations has no impact on any of our major conclusions. All newly-analyzed Swift XRT data are presented in Table \ref{tab:xrays} in Appendix \ref{sec:data}.

\subsection{Optical: Archival spectroscopy and photometry}\label{sec:opt}

Optically selected TDEs typically show a common set of spectral features, including hydrogen and/or helium emission lines \citep{ags+14}. {The relative strengths of these emission lines reveal the ionization state of the stellar debris cloud \citep{rkgr16} and may provide clues to the underlying physics of the disruption; for example, \cite{nlr+22} found that TDEs with only hydrogen in their spectra may be either incomplete stellar disruptions or cases where accretion disk formation is delayed.}

All of the TDEs in our sample have optical spectral classifications previously reported in the literature \citep{fbc+15,Blanchard17,ghw+19,fgv+20,vgh+21,abm+22,hvg+23,yrg+23}. We divide our sample into three spectral types: TDEs with only hydrogen lines in their spectra (hereafter TDE-H), TDEs with both hydrogen and helium lines (TDE-H+He), and TDEs with only helium lines (TDE-He).  However, the relative strengths of these lines vary both across TDEs \citep{ags+14,vgh+21} and as a function of time for specific TDEs (e.g., \citealt{nbb+19}). {For this paper, we follow common practice (e.g., \citealt{vgh+21,hvg+23}) and define each TDE's spectral class based on the lines visible near peak optical light. By this definition, our sample consists of 8 TDE-H, 21 TDE-H+He, and 2 TDE-He. We list the spectral classifications of our TDEs in Table \ref{tab:gold}. While most of these are consistent with the classifications reported in \yvette, who conservatively define a TDE H+He as a TDE that exhibited He lines at \emph{any} phase of its evolution,} three of the TDEs in our sample (ASASSN-14ae, AT2018zr, and AT2018hyz) initially showed only H lines in their spectra, but later developed He II emission at $\sim50-200$ d post-discovery \citep{hpb+14,snl+20,hcr+19}. {While \yvette\ classify these three transients as TDE-H+He, we choose to retain a TDE-H classification for them. This is the best choice to ensure consistency across our sample because 4 of our remaining 5 TDE-H have no published optical spectra at $t>50$ d, so it remains possible that they could have developed He lines at late times as well.} 

We also collected archival optical and ultraviolet photometry to perform more detailed light curve modeling of the TDEs in our sample, as described in Section \ref{sec:mod}. We exclude DES14C1kia from this analysis, as the optical dataset for this TDE remains unpublished. A complete listing of the data used and the original references for each dataset are given in Appendix \ref{sec:data}.

\subsection{Host galaxy observations}\label{sec:host}

Most of our TDEs also have well-studied host galaxies. For our analysis in Section \ref{sec:bhmass}, we use published galaxy stellar masses ($M_{\rm gal}$) calculated via broadband modeling of archival near-IR through UV photometry and published stellar velocity dispersions ($\sigma*$) derived from high-resolution optical spectroscopy \citep{wvj+17,wsv+19,yrg+23}. We also use the optical $u-r$ colors of the TDE host galaxies in our sample as compiled by \cite{fwl+20} and references therein \citep{vgh+21,hgv+21,yrg+23}.

\section{Multi-Wavelength Analysis}\label{sec:mod} 

In this section, we briefly consider basic host galaxy properties of the TDEs in our sample and we compare the optical, UV, and X-ray emission of our TDEs to their radio emission. We model the optical and UV light curves of our TDEs to calculate key parameters of each disruption (e.g., the SMBH mass and the mass of the disrupted star) and search for possible correlations with their radio properties. We then extract the temporal evolution of the mass accretion rate onto the SMBH, $\dot{M}$, and compare the observed radio and X-ray evolution to the timing of potential state transitions in the accretion flow based on our derived $\dot{M}$.  
Finally, we compare the radio and X-ray luminosities of our TDEs to those of AGN, to gain additional insight into $\dot{M}$ and search for possible similarities with the accretion states of those systems.

\subsection{Optical spectroscopic classification, host properties, and SMBH masses}\label{sec:bhmass}

We first search for possible correlations between our radio classifications and the basic optical properties of the TDEs in our sample and their host galaxies. {We find that 7 of the 12 TDEs that exhibit delayed radio emission have an optical spectral classification of TDE-H. For our sample of 31 objects containing 8 TDE-H total, the probability of at least 7 of the 8 TDEs also falling into this group by chance is $p=0.002$. This suggests a correlation between a TDE's optical spectral type and the timing of its radio emission at the $\sim3\sigma$ level. We discuss the implications of this further in Section \ref{sec:disc}.}  

Previous studies of the optical $u-r$ colors of TDE host galaxies have shown that TDE hosts are preferentially located in the ``green valley'' (\citealt{fwl+20} and references therein; \citealt{hgv+21,yrg+23}). We find no obvious correlation between the bulk host galaxy classifications (quiescent vs.\ star forming; red vs.\ green vs.\ blue) and the type of radio emission in our TDE sample (prompt vs.\ delayed vs.\ none/host). We defer further investigation of possible impacts of the host environment on the radio properties of TDEs to future work.

Finally, we compute the SMBH masses for all TDEs in our sample. As SMBH masses previously reported in the literature are derived using several different methods, here we use two independent techniques: a transient-independent method based on observations of the TDE host galaxies ($M_{\rm BH,host}$) 
and a TDE-only method based on fitting the TDE optical and UV light curves ($M_{\rm BH,mosfit}$). 
First, we use archival host galaxy observations to self-consistently recalculate the SMBH masses for all of our TDEs and present them in Table \ref{tab:bhmass}, along with the original measurement used to derive each value. For galaxies with published stellar velocity dispersions, we calculate the black hole mass using the $M_{\rm BH}-\sigma*$ relation from \cite{kh13}. For galaxies that do not have published velocity dispersions but do have published stellar masses, we use the $M_{\rm BH}-M_{\rm gal}$ relation for TDE host galaxies derived by \cite{yrg+23}, which is calibrated against the \cite{kh13} relation.\footnote{OGLE17aaj's host galaxy does not have a published $\sigma*$ measurement or $M_{\rm gal}$ value; we therefore use the SMBH mass estimate from \cite{ghw+19}, $\log (M_{\rm BH}/M_{\odot}) =7.37$.}

TDE light curves have also been used to estimate SMBH masses directly \citep{mgr19,rkp20,sm24,mvn+24}. The masses derived in this way are typically found to be consistent with SMBH masses derived from the host galaxy properties, with some scatter.\footnote{{While a detailed intercomparison of the different light curve models is beyond the scope of this work, we refer the interested reader to Table 7 and the discussion in Section 7 of \cite{hvg+23} for a comparison of the SMBH masses computed using two of these light curve-based models, MOSFiT \citep{mgr19} and TDEMass \citep{rkp20}.}} We choose to model the optical and UV light curves of our TDEs to obtain a second, independent estimate of the SMBH mass for each TDE using {the most widely-used of these three light curve-based models:} the Modular Open Source Fitter for Transients (MOSFiT; \citealt{gnv+18,mgr19}). For our sample, we find that the {host and MOSFiT} methods yield generally consistent results, although on average the MOSFiT-derived masses are slightly higher (Figure \ref{fig:Rph0M*}, left). This is similar to the results obtained by \cite{mgr19} for their TDE sample. For our analysis, we do not make any claims as to which method of deriving the SMBH mass is more accurate; instead, we consider both sets of masses independently whenever possible. We discuss our MOSFiT modeling further in the next two sections. 

\begin{deluxetable*}{lcclc}
\tablenum{2}\label{tab:bhmass}
\tablecaption{SMBH masses for the TDEs in our sample derived from properties of their host galaxies. The $M_{\rm BH,host}$ values have been computed using the $M_{\rm BH}-\sigma*$ relation of \cite{kh13} when possible, or using the $M_{\rm gal}-M_{\rm BH}$ relation from \cite{yrg+23} otherwise. The source used for each galaxy stellar mass or $\sigma*$ measurement is indicated in the Reference column.}
\tablewidth{0pt}
\tablehead{
\colhead{TDE Name} & \colhead{$\log(M_{\rm gal}/M_{\odot})$} & \colhead{$\sigma*$ (km s$^{-1}$)} & \colhead{Reference} & \colhead{$\log(M_{\rm BH,host}/M_{\odot})$} }
\startdata
ASASSN-14ae	& 	--					& $	53	\pm	2	$ & 	\cite{Wevers19}	& $	6.0	\pm	0.3	$ \\
DES14C1kia	& $	10.1					$ & 	--			&	\cite{Wevers19}	& $	6.6 			$ \\
ASASSN-14li	& 	--					& $	81	\pm	2	$ & 	\cite{Wevers19}	& $	6.8	\pm	0.3	$ \\
iPTF15af	& 	--					& $	106	\pm	2	$ & 	\cite{Wevers19}	& $	7.3	\pm	0.3	$ \\
ASASSN-15oi	& 	--					& $	61	\pm	7	$ & 	\cite{Wevers19}	& $	6.2	\pm	0.4	$ \\
iPTF16axa	& 	--					& $	82	\pm	3	$ & 	\cite{Wevers19}	& $	6.8	\pm	0.3	$ \\
PS16dtm	& 	--					& $	45	\pm	3	$ & 	\cite{xbg+11}	& $	5.7	\pm	0.4	$ \\
iPTF16fnl	& 	--					& $	55	\pm	2	$ & 	\cite{Wevers19}	& $	6.0	\pm	0.3	$ \\
OGLE17aaj*	&	--					&	--			&	\cite{ghw+19}	& $	7.37			$ \\
AT2017eqx	& $	9.36	^{+0.1	}_{-0.1	}$ & 	--			&	\cite{nbb+19}	& $	5.3	\pm	0.4	$ \\
AT2018zr	& 	--					& $	49.79	\pm	4.93	$ & 	\cite{hcg+23}	& $	5.8	\pm	0.4	$ \\
AT2018bsi	& 	--					& $	117.54	\pm	8.12	$ & 	\cite{hcg+23}	& $	7.5	\pm	0.3	$ \\
AT2018dyb	& 	--					& $	96	\pm	1	$ & 	\cite{lda+19}	& $	7.1	\pm	0.3	$ \\
AT2018fyk	& 	--					& $	158	\pm	1	$ & 	\cite{wevers20}	& $	8.0	\pm	0.3	$ \\
AT2018hco	& $	10.03	^{+0.12	}_{-0.16	}$ & 	--			&	\cite{hvg+23}	& $	6.5	\pm	0.3	$ \\
AT2018hyz	& 	--					& $	66.62	\pm	3.12	$ & 	\cite{hcg+23}	& $	6.4	\pm	0.3	$ \\
AT2018lna	& 	--					& $	36.43	\pm	4.52	$ & 	\cite{hcg+23}	& $	5.2	\pm	0.4	$ \\
ASASSN-19bt	& $	10.04	^{+0.34	}_{-0.04	}$ & 	--			&	\cite{hva+19}	& $	6.5	\pm	0.4	$ \\
AT2019azh	& 	--					& $	68.01	\pm	2.02	$ & 	\cite{hcg+23}	& $	6.4	\pm	0.3	$ \\
AT2019dsg	& 	--					& $	86.89	\pm	3.92	$ & 	\cite{yrg+23}	& $	6.9	\pm	0.3	$ \\
AT2019ehz	& 	--					& $	46.65	\pm	11.83	$ & 	\cite{hcg+23}	& $	5.7	\pm	0.6	$ \\
AT2019eve	& $	9.26	^{+0.11	}_{-0.17	}$ & 	--			&	\cite{vgh+21}	& $	5.1	\pm	0.4	$ \\
AT2019qiz	& 	--					& $	71.85	\pm	1.93	$ & 	\cite{hcg+23}	& $	6.5	\pm	0.3	$ \\
AT2019teq	& $	9.95	^{+0.07	}_{-0.11	}$ & 	--			&	\cite{hvg+23}	& $	6.3	\pm	0.3	$ \\
AT2020pj	& $	10.01	^{+0.07	}_{-0.08	}$ & 	--			&	\cite{yrg+23}	& $	6.4	\pm	0.3	$ \\
AT2020mot	& 	--					& $	76.61	\pm	5.33	$ & 	\cite{yrg+23}	& $	6.7	\pm	0.3	$ \\
AT2020neh	& 	--					& $	40	\pm	6	$ & 	\cite{abm+22}	& $	5.4	\pm	0.5	$ \\
AT2020nov	& -- & 	$127 \pm 28$	&	
\cite{efr+24} & $	7.6	\pm	0.5	$ \\
AT2020opy	& $	10.01	^{+0.13	}_{-0.14	}$ & 	--			&	\cite{hvg+23}	& $	6.4	\pm	0.3	$ \\
AT2020wey	& 	--					& $	53.54	\pm	4.75	$ & 	\cite{hcg+23}	& $	6.0	\pm	0.4	$ \\
AT2020vwl	& 	--					& $	48.49	\pm	2.00	$ & 	\cite{yrg+23}	& $	5.8	\pm	0.4	$ \\
\enddata
\smallskip
*OGLE17aaj's host galaxy does not have a published stellar mass or velocity dispersion; we therefore assume the value of $\log \left( M_{\rm BH} / M_{\odot} \right) \sim7.37$ computed by \cite{ghw+19}.
\end{deluxetable*}

\subsection{{MOSFiT modeling: physical parameters}\label{sec:mosfit}}

We fit the optical and UV photometry for the TDEs in our sample with MOSFiT \citep{gnv+18}, using version 1.1.9 of the TDE module first developed by \cite{mgr19} following \cite{gr13,gmr14}. MOSFiT assumes that mass falling onto the SMBH accretes promptly, such that the observed optical luminosity is directly proportional to {both} the mass fallback rate and the accretion rate {and we can define} $L=\epsilon\dot{M}c^2$ (where $\epsilon$ is the radiative efficiency).\footnote{A possible viscous delay is included in the model, but when applied to our observations this delay is generally found to be very short (consistent with zero in most cases), in line with previous work \citep{mgr19,mr21,nlr+22}.} This corresponds to Case 1 as defined in the introduction. This assumption likely breaks down at very late times, when TDEs' optical/UV emission often exhibits a plateau phase that is not well-captured by MOSFiT but can be better modeled by a bare accretion disk \citep{vsm+19, mnig24, mvn+24}. For the TDEs in our sample that show optical/UV plateaus, we therefore only include photometry prior to the onset of the plateau phase in our modeling. The original references for the included data, {together with the photometry cutoff times where applicable,} are given in Table \ref{tab:data} in Appendix \ref{sec:data}. 

The MOSFiT TDE model has nine free physical parameters \citep{mgr19,nlr+22}, of which eight are of potential interest for discriminating among competing TDE emission models: the viscous delay time $t_v$, the disruption impact parameter $b$, the SMBH mass $M_{\rm BH,mosfit}$, the column density along the line of sight in the host galaxy $n_{\rm H,host}$, the normalization of the photospheric radius $R_{\rm ph,0}$ at the time when the observed bolometric luminosity of the TDE reaches $L_{\rm Edd}$, the corresponding power law exponent $l_{\rm ph}$ that relates the photospheric radius to the instantaneous luminosity as $R_{\rm ph}\propto R_{\rm ph,0}(L/L_{\rm Edd})^{l_{\rm ph}}$, the mass of the disrupted star $M_*$, and the efficiency with which accreting mass is converted to radiation, $\epsilon$. The remaining parameter is the time of first fallback relative to the first data point in the light curve, $t_0$. Although not directly physically relevant, $t_0$ can be used in combination with the light curves to calculate the date of disruption, MJD$_0$, and the number of rest-frame days between disruption and peak optical/UV light, $t_{\rm rise}$ (which has been suggested to correlate with other physical properties of TDEs; e.g., \citealt{r88,vgh+21}). We derive two additional parameters of possible physical interest from our MOSFiT solutions: $\dot{M}_{\rm max}$, the maximum accretion rate onto the SMBH during each TDE, and $L_{\rm max}$, the maximum bolometric luminosity during the TDE. For our analysis, we normalize $\dot{M}_{\rm max}$ by $\dot{M}_{\rm Edd}$ (as defined in Section \ref{sec:mosfit2}) to remove the dependence on the efficiency. 

{Using} the same setup, priors, and model assumptions as those described in \cite{nlr+22},   
we are able to find a converged model solution for all TDEs in our sample except for DES141Ckia (which has no published photometry) and AT2018fyk (whose double-peaked light curve is not well-modeled by MOSFiT's one-zone model assumptions). The best-fit physically relevant parameters and $1\sigma$ confidence intervals for our fits are given in Table \ref{tab:mosfit}.  We note that our results for the TDEs also studied by \cite{nlr+22} are similar to those reported in that work.

\begin{figure}
\centering
    \includegraphics[height=2.75in]{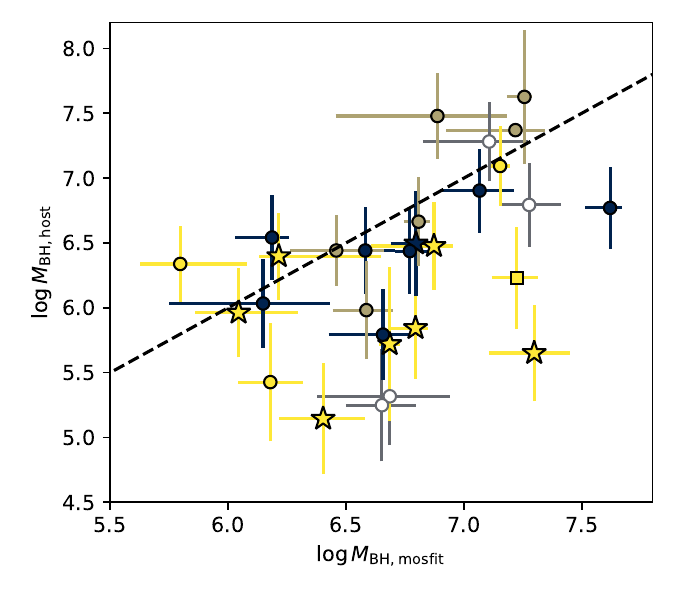}\includegraphics[height=2.75in]{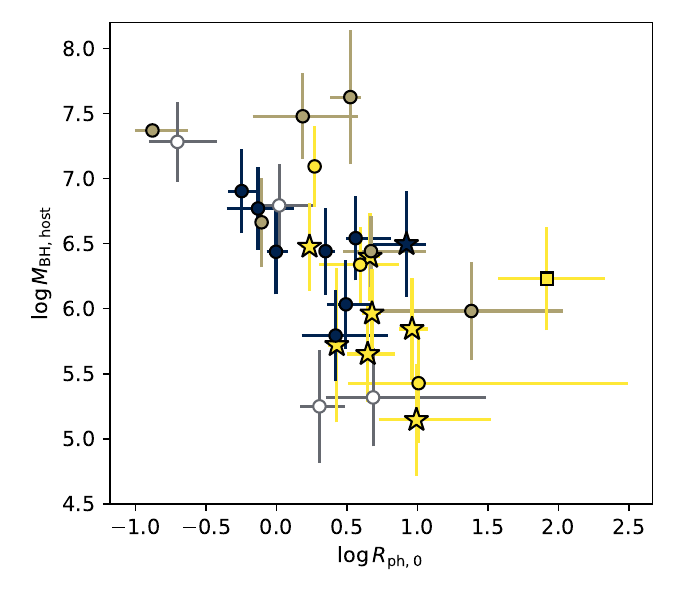}
    
    \caption{\label{fig:Rph0M*} Left: The host-derived BH masses $M_{\rm BH,host}$ compared to the MOSFiT-derived BH masses $M_{\rm BH,mosfit}$ for our sample. The dashed line shows the 1:1 relation. The marker color indicates the timing of radio emission for each TDE and the symbol indicates its optical spectral type. Right: $M_{\rm BH,host}$ compared to the MOSFiT-derived normalized photospheric radii $R_{\rm ph,0}$. Delayed radio TDEs (yellow points) exhibit larger $R_{\rm ph,0}$ than Prompt radio TDEs (dark blue points) and are more likely to have optical spectral type TDE-H (stars).}
\end{figure}

We first use these results to search for possible correlations between the radio properties of our TDEs and other physical parameters of the disruption. We divide our sample into three groups for this analysis:  those with radio type ``Prompt'' in Table \ref{tab:gold} (i.e., TDEs that show early/prompt radio emission), ``Delayed'' TDEs in Table \ref{tab:gold} (those that show late/delayed radio emission), and ``None'' or ``Host'' TDEs in Table \ref{tab:gold} (those that show no radio emission or only host or faint ambiguous radio emission). We also considered the None and Host TDEs as two independent categories, but this does not significantly change our results. We then performed an Anderson-Darling (AD) test\footnote{The Anderson-Darling test is similar to the Kolmogorov–Smirnov (KS) test used by some previous authors (e.g., \citealt{nlr+22}), but it has been shown to have added sensitivity to the tails of the distributions in datasets with large numbers of free parameters \citep{hphw09}. We note that a KS test applied to our data gives similar results.} for each of the parameters listed in Table \ref{tab:mosfit} plus $M_{\rm BH,host}$ (the SMBH masses derived from host galaxy properties presented in Table \ref{tab:bhmass}). 
In almost all cases, we find no obvious differences in the parameter distributions between any two of the three radio classes (Table \ref{tab:ad}). The one exception is the photospheric radius normalization parameter $R_{\rm ph,0}$, which is larger for the delayed radio TDEs than for the prompt radio TDEs ($p=0.026$). There is also a marginal suggestion ($p\sim0.05$) that the TDEs with delayed radio emission have lower SMBH masses than either the TDEs with prompt radio emission or than the host+none group -- but only when $M_{\rm BH,host}$ is used; no significant differences are found between the $M_{\rm BH,mosfit}$ distributions for these groups. We plot the distributions of $R_{\rm ph,0}$, $M_{\rm BH,host}$, and $M_{\rm BH,mosfit}$ in Figure \ref{fig:Rph0M*}. While achieving three $p$ values $\lesssim0.05$ out of 36 statistical tests performed is not a highly significant result, in Section \ref{sec:disc} we briefly explore the possible implications if these correlations are verified with a larger TDE sample in future work. We emphasize that null results in our current sample do not necessarily imply that correlations between the radio properties and the various parameters tested are not present, given the limitations of existing data.

\movetabledown=1.8in
\begin{rotatetable}
\begin{deluxetable}{lrrrrrrrrrr}
\tablenum{3}\label{tab:mosfit}
\tablecaption{MOSFiT Parameters}
\tablewidth{0pt}
\tablehead{
\colhead{TDE Name} & \colhead{$t_v$ (d)}  & \colhead{$b$} & \colhead{$\log(M_{\rm BH,mosfit}/$} &  \colhead{$\log{(n_{\rm H,host}/}$} & \colhead{$R_{\rm ph,0}$} & \colhead{$M_*$} & \colhead{$\epsilon$} &  \colhead{$t_{\rm rise}$ (d)} & \colhead{$\dot{M}_{\rm max}/$} & \colhead{$L_{\rm max}$} \\ 
\nocolhead{Name} & \colhead{(rest-frame)} & \nocolhead{Number} &  \colhead{$M_{\odot})$} & \colhead{cm$^{-2}$)} & \nocolhead{(km)} & \colhead{($M_{\odot}$)} & \nocolhead{Number} & \colhead{(rest-frame)} & \colhead{$\dot{M}_{\rm Edd}$} & \colhead{($10^{44}$ erg s$^{-1}$)} } 
\startdata
ASASSN-14ae & $0.14^{+2.35}_{-0.13}$ & $0.30^{+0.12}_{-0.10}$ & $6.0^{+0.3}_{-0.2}$ & $18.1^{+1.5}_{-1.3}$ & $4.8^{+0.8}_{-0.7}$ & $1.0^{+0.2}_{-0.2}$ & $0.009^{+0.019}_{-0.005}$ & $27^{+9}_{-11}$ & $	0.7^{+1.0}_{-0.4}$ & $0.65^{+0.07}_{-0.13}$ \\
ASASSN-14li	 & $ 	0.12	^{+	6.48	}_{-	0.12	}$ & $	0.64	^{+	0.21	}_{-	0.08	}$ & $	7.62	^{+	0.05	}_{-	0.11	}$ & $	21.11	^{+	0.02	}_{-	0.02	}$ & $	0.7	^{+	0.6	}_{-	0.3	}$ & $	0.8	^{+	0.2	}_{-	0.2	}$ & $	0.23	^{+	0.12	}_{-	0.13	}$ & $	132	^{+	10	}_{-	16	}$ & $	0.5	^{+	0.3	}_{-	0.2	}$ & $	21	^{+	9	}_{-	6	} $ \\
iPTF15af	 & $ 	0.4	^{+	18.3	}_{-	0.4	}$ & $	0.7	^{+	0.6	}_{-	0.3	}$ & $	7.1	^{+	0.2	}_{-	0.3	}$ & $	21.00	^{+	0.05	}_{-	0.06	}$ & $	0.20	^{+	0.18	}_{-	0.07	}$ & $	1.4	^{+	0.9	}_{-	0.8	}$ & $	0.07	^{+	0.16	}_{-	0.06	}$ & $	28	^{+	12	}_{-	16	}$ & $	1.3	^{+	3.8	}_{-	1.1	}$ & $	8	^{+	5	}_{-	7	} $ \\
ASASSN-15oi	 & $ 	0.05	^{+	1.09	}_{-	0.05	}$ & $	0.94	^{+	0.02	}_{-	0.06	}$ & $	7.22	^{+	0.09	}_{-	0.11	}$ & $	17.6	^{+	1.1	}_{-	0.9	}$ & $	83	^{+	131	}_{-	45	}$ & $	1.00	^{+	0.05	}_{-	0.05	}$ & $	0.031	^{+	0.006	}_{-	0.006	}$ & $	131	^{+	15	}_{-	8	}$ & $	0.6	^{+	0.2	}_{-	0.2	}$ & $	8.5	^{+	0.8	}_{-	1.1	} $ \\
iPTF16axa	 & $ 	0.2	^{+	3.2	}_{-	0.2	}$ & $	0.93	^{+	0.11	}_{-	0.13	}$ & $	7.28	^{+	0.13	}_{-	0.12	}$ & $	21.17	^{+	0.02	}_{-	0.03	}$ & $	1.0	^{+	0.9	}_{-	0.4	}$ & $	0.5	^{+	0.4	}_{-	0.2	}$ & $	0.24	^{+	0.11	}_{-	0.10	}$ & $	142	^{+	29	}_{-	31	}$ & $	1.3	^{+	1.0	}_{-	0.6	}$ & $	15	^{+	4	}_{-	5	} $ \\
PS16dtm	 & $ 	84	^{+	26	}_{-	25	}$ & $	0.63	^{+	0.14	}_{-	0.19	}$ & $	7.3	^{+	0.2	}_{-	0.2	}$ & $	17.3	^{+	1.3	}_{-	0.8	}$ & $	4	^{+	2	}_{-	1	}$ & $	0.5	^{+	0.3	}_{-	0.1	}$ & $	0.05	^{+	0.05	}_{-	0.02	}$ & $	104	^{+	7	}_{-	11	}$ & $	0.2	^{+	0.2	}_{-	0.1	}$ & $	2.51	^{+	0.08	}_{-	0.08	} $ \\
iPTF16fnl	 & $ 	0.07	^{+	0.51	}_{-	0.06	}$ & $	1.04	^{+	0.05	}_{-	0.57	}$ & $	6.1	^{+	0.3	}_{-	0.4	}$ & $	20.8	^{+	0.1	}_{-	3.0	}$ & $	3.1	^{+	1.5	}_{-	0.8	}$ & $	0.105	^{+	1.113	}_{-	0.008	}$ & $	0.005	^{+	0.003	}_{-	0.001	}$ & $	17	^{+	4	}_{-	9	}$ & $	0.4	^{+	0.8	}_{-	0.1	}$ & $	0.5	^{+	0.3	}_{-	0.6	} $ \\
OGLE17aaj	 & $ 	5	^{+	25	}_{-	5	}$ & $	0.98	^{+	0.12	}_{-	0.16	}$ & $	7.22	^{+	0.13	}_{-	0.29	}$ & $	19	^{+	2	}_{-	1	}$ & $	0.13	^{+	0.10	}_{-	0.03	}$ & $	0.7	^{+	1.1	}_{-	0.4	}$ & $	0.03	^{+	0.07	}_{-	0.02	}$ & $	14	^{+	3	}_{-	4	}$ & $	0.5	^{+	0.9	}_{-	0.3	}$ & $	6.4	^{+	3.2	}_{-	5.0	} $ \\
AT2017eqx	 & $ 	0.04	^{+	0.91	}_{-	0.04	}$ & $	1.0	^{+	0.1	}_{-	0.2	}$ & $	6.7	^{+	0.3	}_{-	0.3	}$ & $	21.0	^{+	0.1	}_{-	2.9	}$ & $	5	^{+	26	}_{-	3	}$ & $	0.2	^{+	0.3	}_{-	0.1	}$ & $	0.02	^{+	0.05	}_{-	0.01	}$ & $	39	^{+	13	}_{-	17	}$ & $	0.3	^{+	0.8	}_{-	0.2	}$ & $	1.9	^{+	1.5	}_{-	2.6	} $ \\
AT2018zr	 & $ 	0.01	^{+	0.31	}_{-	0.01	}$ & $	0.87	^{+	0.07	}_{-	0.07	}$ & $	6.79	^{+	0.05	}_{-	0.11	}$ & $	20.92	^{+	0.05	}_{-	0.07	}$ & $	9	^{+	3	}_{-	2	}$ & $	0.11	^{+	0.07	}_{-	0.04	}$ & $	0.04	^{+	0.02	}_{-	0.02	}$ & $	8.0	^{+	0.7	}_{-	0.8	}$ & $	0.2	^{+	0.1	}_{-	0.0	}$ & $	1.5	^{+	0.2	}_{-	0.3	} $ \\
AT2018bsi	 & $ 	0.1	^{+	3.1	}_{-	0.1	}$ & $	0.9	^{+	0.3	}_{-	0.2	}$ & $	6.9	^{+	0.3	}_{-	0.4	}$ & $	20.6	^{+	0.2	}_{-	0.8	}$ & $	1.5	^{+	2.2	}_{-	0.9	}$ & $	0.7	^{+	0.6	}_{-	0.2	}$ & $	0.011	^{+	0.033	}_{-	0.005	}$ & $	78	^{+	32	}_{-	58	}$ & $	0.3	^{+	0.8	}_{-	0.2	}$ & $	2.1	^{+	1.0	}_{-	2.3	} $ \\
AT2018dyb	 & $ 	0.06	^{+	0.53	}_{-	0.06	}$ & $	0.90	^{+	0.08	}_{-	0.03	}$ & $	7.15	^{+	0.04	}_{-	0.04	}$ & $	17.2	^{+	0.7	}_{-	0.7	}$ & $	1.9	^{+	0.2	}_{-	0.2	}$ & $	0.33	^{+	0.04	}_{-	0.09	}$ & $	0.23	^{+	0.10	}_{-	0.04	}$ & $	25	^{+	2	}_{-	4	}$ & $	1.1	^{+	0.2	}_{-	0.2	}$ & $	10.8	^{+	0.6	}_{-	0.7	} $ \\
AT2018hco	 & $ 	0.1	^{+	2.4	}_{-	0.1	}$ & $	0.77	^{+	0.10	}_{-	0.10	}$ & $	6.87	^{+	0.08	}_{-	0.27	}$ & $	19	^{+	2	}_{-	2	}$ & $	1.7	^{+	0.2	}_{-	0.2	}$ & $	0.12	^{+	0.10	}_{-	0.05	}$ & $	0.08	^{+	0.05	}_{-	0.04	}$ & $	8	^{+	1	}_{-	2	}$ & $	0.32	^{+	0.27	}_{-	0.09	}$ & $	2.4	^{+	0.3	}_{-	0.5	} $ \\
AT2018hyz	 & $ 	0.10	^{+	1.13	}_{-	0.09	}$ & $	0.75	^{+	0.04	}_{-	0.23	}$ & $	6.21	^{+	0.43	}_{-	0.08	}$ & $	19	^{+	1	}_{-	2	}$ & $	4.6	^{+	0.8	}_{-	0.4	}$ & $	1.2	^{+	0.1	}_{-	0.8	}$ & $	0.0035	^{+	0.0089	}_{-	0.0007	}$ & $	7	^{+	2	}_{-	3	}$ & $	1.6	^{+	0.6	}_{-	1.4	}$ & $	1.31	^{+	0.07	}_{-	0.09	} $ \\
AT2018lna	 & $ 	0.10	^{+	1.90	}_{-	0.09	}$ & $	0.97	^{+	0.12	}_{-	0.19	}$ & $	6.65	^{+	0.15	}_{-	0.15	}$ & $	18	^{+	2	}_{-	2	}$ & $	2.0	^{+	1.1	}_{-	0.5	}$ & $	0.6	^{+	0.3	}_{-	0.2	}$ & $	0.007	^{+	0.003	}_{-	0.003	}$ & $	15	^{+	5	}_{-	7	}$ & $	0.4	^{+	0.3	}_{-	0.1	}$ & $	1.7	^{+	0.2	}_{-	0.4	} $ \\
ASASSN-19bt	 & $ 	1.6	^{+	1.2	}_{-	0.6	}$ & $	0.89	^{+	0.12	}_{-	0.14	}$ & $	6.80	^{+	0.05	}_{-	0.11	}$ & $	20.79	^{+	0.03	}_{-	0.10	}$ & $	8	^{+	3	}_{-	4	}$ & $	0.7	^{+	1.0	}_{-	0.1	}$ & $	0.009	^{+	0.004	}_{-	0.005	}$ & $	9.5	^{+	0.9	}_{-	1.2	}$ & $	0.4	^{+	0.2	}_{-	0.1	}$ & $	2.5	^{+	0.4	}_{-	0.2	} $ \\
AT2019azh	 & $ 	0.06	^{+	0.63	}_{-	0.06	}$ & $	0.32	^{+	0.24	}_{-	0.05	}$ & $	6.77	^{+	0.09	}_{-	0.06	}$ & $	19.2	^{+	0.9	}_{-	2.1	}$ & $	1.0	^{+	0.2	}_{-	0.1	}$ & $	0.6	^{+	0.2	}_{-	0.2	}$ & $	0.13	^{+	0.09	}_{-	0.04	}$ & $	9	^{+	1	}_{-	2	}$ & $	0.9	^{+	0.5	}_{-	0.2	}$ & $	3.7	^{+	0.3	}_{-	0.9	} $ \\
AT2019dsg	 & $ 	0.1	^{+	2.5	}_{-	0.1	}$ & $	0.89	^{+	0.18	}_{-	0.13	}$ & $	7.07	^{+	0.14	}_{-	0.17	}$ & $	19	^{+	1	}_{-	2	}$ & $	0.6	^{+	0.2	}_{-	0.1	}$ & $	0.5	^{+	0.3	}_{-	0.2	}$ & $	0.08	^{+	0.13	}_{-	0.05	}$ & $	36	^{+	10	}_{-	12	}$ & $	0.9	^{+	0.7	}_{-	0.3	}$ & $	7.7	^{+	1.7	}_{-	2.4	} $ \\
AT2019ehz	 & $ 	0.2	^{+	1.3	}_{-	0.2	}$ & $	0.46	^{+	0.10	}_{-	0.07	}$ & $	6.69	^{+	0.04	}_{-	0.04	}$ & $	20.92	^{+	0.04	}_{-	0.05	}$ & $	2.7	^{+	0.5	}_{-	0.4	}$ & $	0.14	^{+	0.02	}_{-	0.03	}$ & $	0.28	^{+	0.07	}_{-	0.09	}$ & $	10	^{+	1	}_{-	2	}$ & $	1.1	^{+	0.5	}_{-	0.4	}$ & $	3.6	^{+	0.7	}_{-	0.5	} $ \\
AT2019eve	 & $ 	0.04	^{+	0.67	}_{-	0.03	}$ & $	0.14	^{+	0.13	}_{-	0.06	}$ & $	6.4	^{+	0.2	}_{-	0.2	}$ & $	19	^{+	2	}_{-	2	}$ & $	10	^{+	24	}_{-	5	}$ & $	0.5	^{+	0.4	}_{-	0.2	}$ & $	0.02	^{+	0.04	}_{-	0.02	}$ & $	16	^{+	4	}_{-	4	}$ & $	0.03	^{+	0.03	}_{-	0.01	}$ & $	0.10	^{+	0.01	}_{-	0.04	} $ \\
AT2019qiz	 & $ 	5	^{+	1	}_{-	4	}$ & $	0.40	^{+	0.07	}_{-	0.26	}$ & $	6.19	^{+	0.07	}_{-	0.16	}$ & $	20.6	^{+	0.2	}_{-	0.7	}$ & $	3.6	^{+	2.9	}_{-	0.5	}$ & $	0.9	^{+	2.6	}_{-	0.2	}$ & $	0.004	^{+	0.002	}_{-	0.002	}$ & $	7	^{+	1	}_{-	3	}$ & $	0.22	^{+	0.09	}_{-	0.08	}$ & $	0.3	^{+	0.1	}_{-	0.2	} $ \\
AT2019teq	 & $ 	16	^{+	13	}_{-	16	}$ & $	1.1	^{+	0.4	}_{-	0.3	}$ & $	5.8	^{+	0.3	}_{-	0.2	}$ & $	20.93	^{+	0.13	}_{-	0.12	}$ & $	4	^{+	3	}_{-	2	}$ & $	0.7	^{+	0.6	}_{-	0.4	}$ & $	0.006	^{+	0.012	}_{-	0.004	}$ & $	21	^{+	8	}_{-	18	}$ & $	5	^{+	8	}_{-	3	}$ & $	0.6	^{+	0.1	}_{-	0.5	} $ \\
AT2020pj	 & $ 	0.07	^{+	0.94	}_{-	0.06	}$ & $	0.8	^{+	0.2	}_{-	0.2	}$ & $	6.5	^{+	0.2	}_{-	0.2	}$ & $	21.21	^{+	0.06	}_{-	0.07	}$ & $	5	^{+	7	}_{-	2	}$ & $	0.4	^{+	0.2	}_{-	0.2	}$ & $	0.010	^{+	0.014	}_{-	0.006	}$ & $	17	^{+	5	}_{-	9	}$ & $	0.5	^{+	1.3	}_{-	0.3	}$ & $	1.4	^{+	0.6	}_{-	0.7	} $ \\
AT2020mot	 & $ 	15	^{+	2	}_{-	3	}$ & $	0.13	^{+	0.04	}_{-	0.04	}$ & $	6.81	^{+	0.05	}_{-	0.06	}$ & $	20.76	^{+	0.07	}_{-	0.12	}$ & $	0.78	^{+	0.10	}_{-	0.06	}$ & $	11	^{+	4	}_{-	4	}$ & $	0.05	^{+	0.04	}_{-	0.03	}$ & $	20	^{+	2	}_{-	2	}$ & $	0.4	^{+	0.1	}_{-	0.1	}$ & $	2.1	^{+	0.5	}_{-	0.4	} $ \\
AT2020neh	 & $ 	0.05	^{+	0.40	}_{-	0.05	}$ & $	0.3	^{+	0.4	}_{-	0.1	}$ & $	6.18	^{+	0.14	}_{-	0.14	}$ & $	18	^{+	2	}_{-	1	}$ & $	10	^{+	300	}_{-	7	}$ & $	0.6	^{+	0.4	}_{-	0.2	}$ & $	0.02	^{+	0.10	}_{-	0.02	}$ & $	8	^{+	3	}_{-	5	}$ & $	0.8	^{+	3.7	}_{-	0.7	}$ & $	0.6	^{+	0.4	}_{-	1.2	} $ \\
AT2020nov	 & $ 	1.0	^{+	3.1	}_{-	0.8	}$ & $	0.99	^{+	0.04	}_{-	0.08	}$ & $	7.26	^{+	0.03	}_{-	0.07	}$ & $	18	^{+	2	}_{-	2	}$ & $	3.3	^{+	0.6	}_{-	0.9	}$ & $	0.31	^{+	0.13	}_{-	0.04	}$ & $	0.010	^{+	0.002	}_{-	0.002	}$ & $	16	^{+	2	}_{-	2	}$ & $	0.034	^{+	0.050	}_{-	0.004	}$ & $	0.85	^{+	0.05	}_{-	0.58	} $ \\
AT2020opy	 & $ 	0.4	^{+	3.6	}_{-	0.4	}$ & $	0.97	^{+	0.10	}_{-	0.10	}$ & $	6.58	^{+	0.13	}_{-	0.15	}$ & $	18.3	^{+	0.9	}_{-	1.3	}$ & $	2.2	^{+	0.4	}_{-	0.2	}$ & $	1.0	^{+	0.4	}_{-	0.2	}$ & $	0.0038	^{+	0.0022	}_{-	0.0008	}$ & $	16	^{+	4	}_{-	8	}$ & $	0.6	^{+	0.4	}_{-	0.2	}$ & $	2.0	^{+	0.1	}_{-	0.1	} $ \\
AT2020wey	 & $ 	0.1	^{+	6.1	}_{-	0.1	}$ & $	0.8	^{+	0.2	}_{-	0.2	}$ & $	6.59	^{+	0.11	}_{-	0.14	}$ & $	20.74	^{+	0.08	}_{-	0.14	}$ & $	24	^{+	84	}_{-	19	}$ & $	0.6	^{+	0.3	}_{-	0.1	}$ & $	0.006	^{+	0.003	}_{-	0.002	}$ & $	19	^{+	3	}_{-	3	}$ & $	0.4	^{+	0.5	}_{-	0.2	}$ & $	1.4	^{+	0.5	}_{-	0.6	} $ \\
AT2020vwl	 & $0.1^{+2.1	}_{-	0.1	}$ & $	0.18	^{+	0.05	}_{-	0.05	}$ & $	6.7	^{+	0.2	}_{-	0.2	}$ & $	20.1	^{+	0.7	}_{-	3.1	}$ & $	3	^{+	4	}_{-	1	}$ & $	1.1	^{+	1.8	}_{-	0.3	}$ & $	0.03	^{+	0.04	}_{-	0.02	}$ & $	90	^{+	25	}_{-	29	}$ & $	0.07	^{+	0.20	}_{-	0.04	}$ & $0.4	^{+	0.1	}_{-	0.7	} $ \\
\enddata
\end{deluxetable}
\end{rotatetable} 

\begin{table}
\tablenum{4}
\caption{Anderson-Darling Test Results. The numbers give the probabilities that the underlying parameter distributions are the same for the two sub-samples being compared in each column. Entries for which $p\leq0.05$ are in {bold}.} 
\label{tab:ad}
\begin{tabular}{lccc}
\toprule
 & Early vs. Late & No Detection + Host vs. Late & No Detection + Host vs. Early \\
\midrule
$M_{\rm BH,host}$ & \textbf{0.050} & \textbf{0.049} & 0.227 \\
$\log M_{\rm BH,mosfit}$ & $>0.25$ & 0.199 & $>0.25$ \\
$t_{\rm viscous}$ & 0.188 & $>0.25$ & $>0.25$ \\
$b$ & $>0.25$ & 0.196 & $>0.25$ \\
$\log n_{\rm H,host}$ & 0.132 & 0.074 & $>0.25$ \\
$R_{\rm ph,0}$ & \textbf{0.026} & 0.056 & $>0.25$ \\
$M_*$  & $>0.25$ & $>0.25$ & $>0.25$ \\
$\epsilon$ & $>0.25$ & $>0.25$ & $>0.25$ \\
$l_{\rm photo}$ & $>0.25$ & $>0.25$ & $>0.25$ \\
$t_{\rm rise}$ & $>0.25$ & 0.151 & $>0.25$ \\
$\dot{M}_{\rm max}$ & $>0.25$ & $>0.25$ & $>0.25$ \\
$L_{\rm max}$ & $>0.25$ & $>0.25$ & $>0.25$ \\
\bottomrule
\end{tabular}
\end{table}

\begin{figure}
\centering
    \includegraphics[width=.5\textwidth]{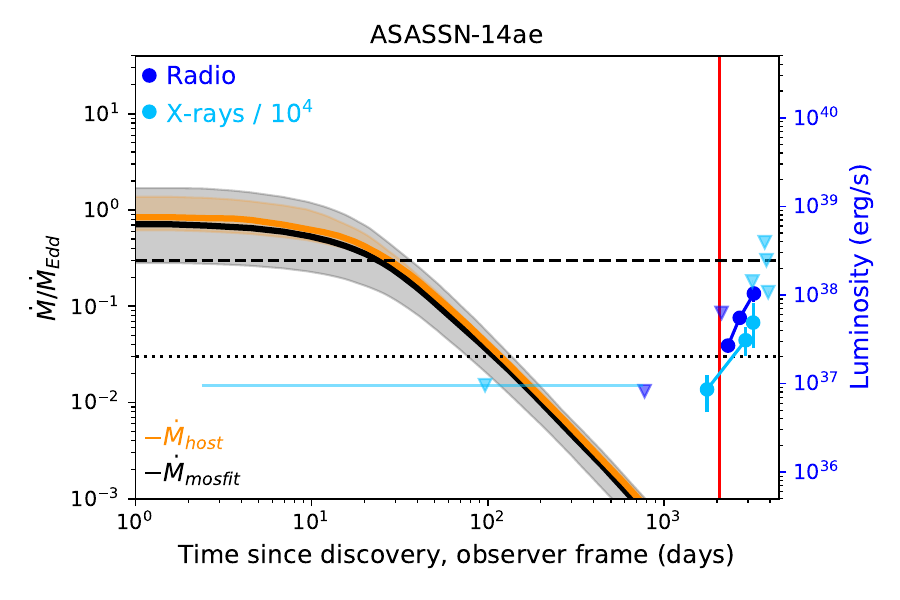}\includegraphics[width=.5\textwidth]{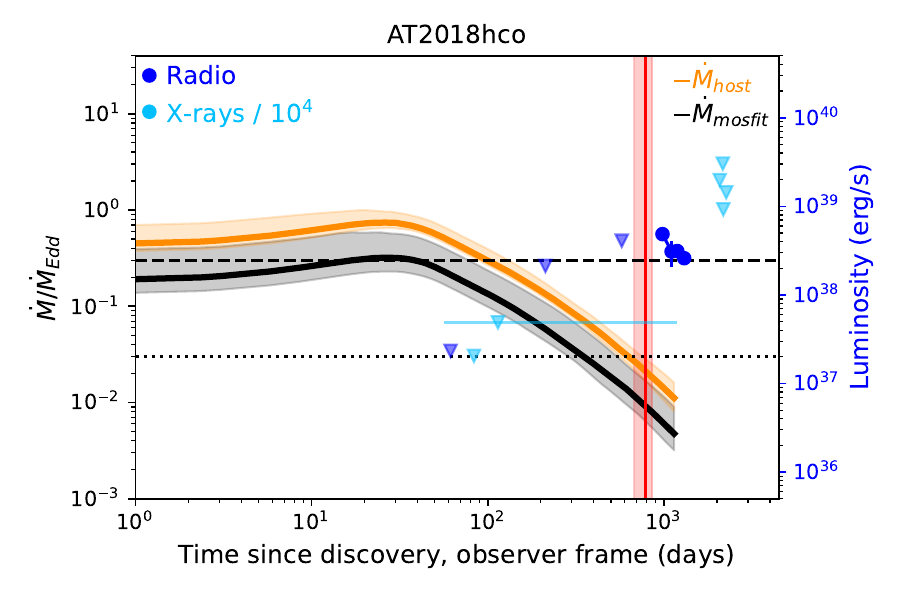}
    \includegraphics[width=.5\textwidth]{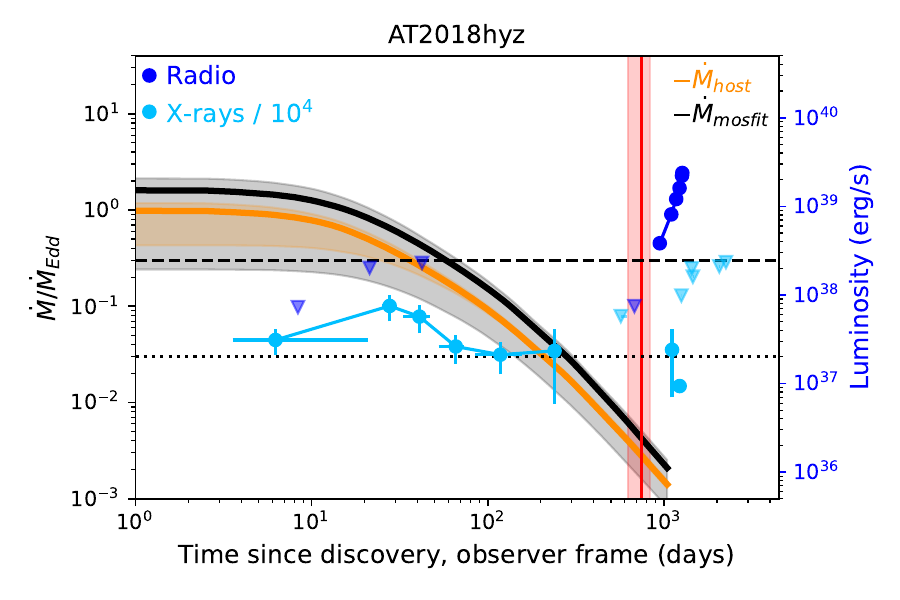}\includegraphics[width=.5\textwidth]{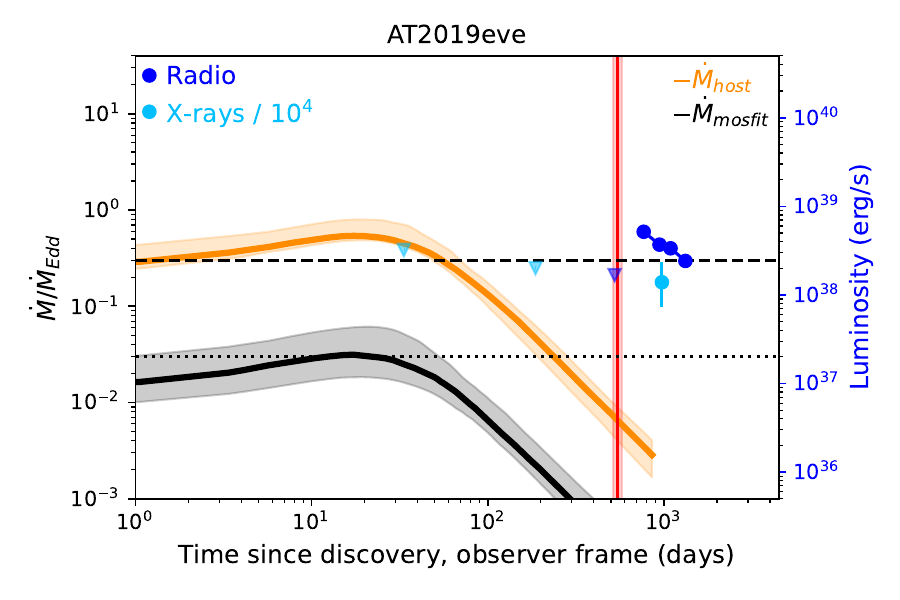} 

    \includegraphics[width=.5\textwidth]{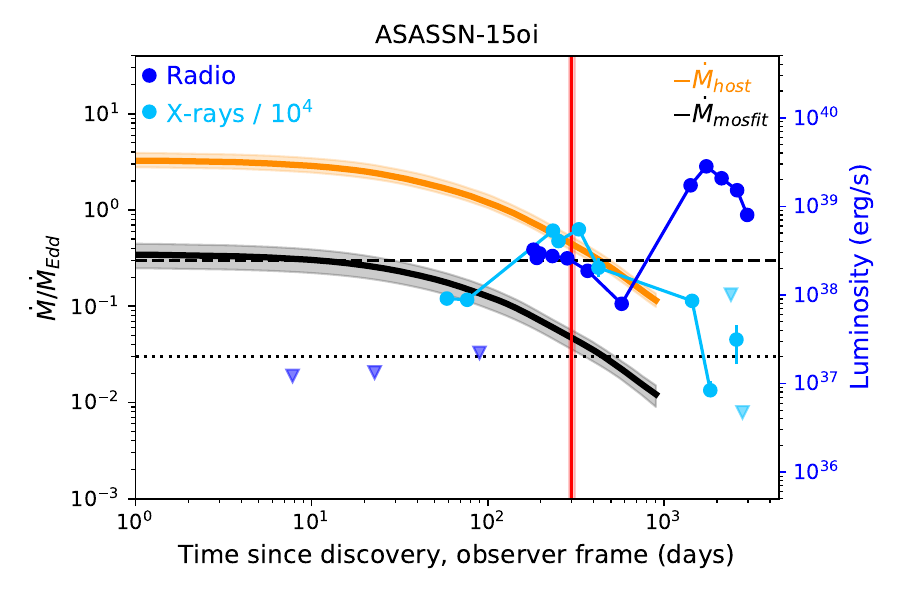}\includegraphics[width=.48\textwidth]{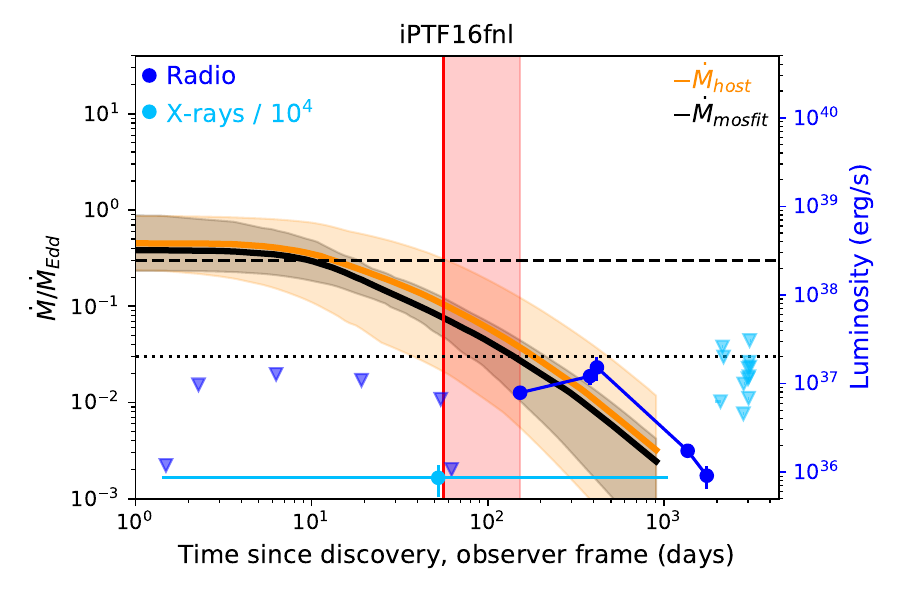} 
    
    \caption{\label{fig:mdot}$\dot{M}$ curves from MOSFiT (solid lines; shaded region shows $1\sigma$ uncertainties) together with the radio (dark blue points) and X-ray (light blue points) light curves of TDEs where the approximate launch time of the outflow can be calculated from analysis of the full multi-frequency radio dataset. We normalize $\dot{M}$ by $\dot{M}_{\rm Edd}$ using both $M_{\rm BH,host}$ (orange) and $M_{\rm BH,mosfit}$ (black). Horizontal lines show the values of $\dot{M}$ corresponding to possible state transitions in the accretion flow; accretion-driven outflows are most likely for $\dot{M}\gtrsim0.3\,\dot{M}_{\rm Edd}$ or $\dot{M} \lesssim0.03\,\dot{M}_{\rm Edd}$. The vertical red lines (shaded bars) show the estimated launch times (and associated uncertainties) of the late-rising radio outflows as described in the text.}
\end{figure}

\begin{figure}
\centering
    \includegraphics[width=.5\textwidth]{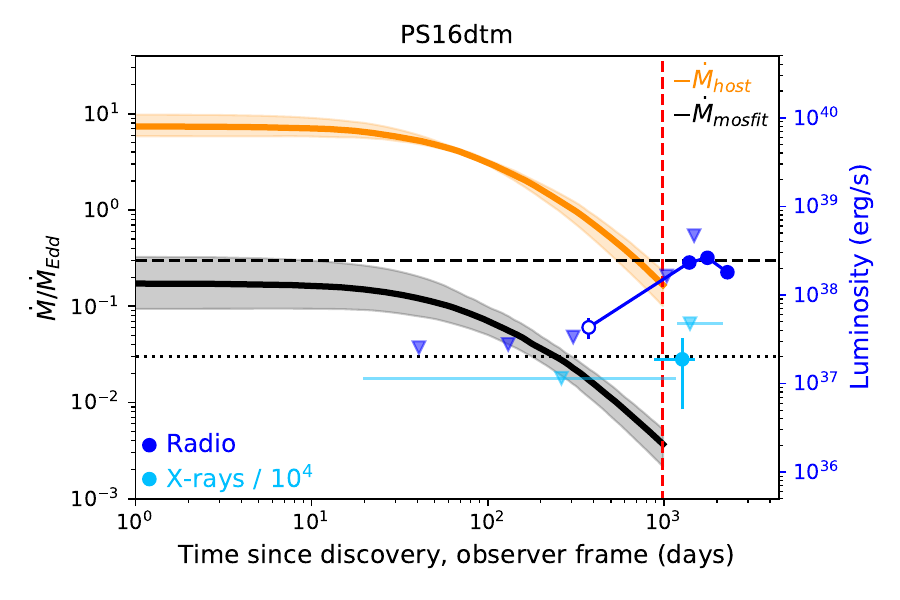}\includegraphics[width=.48\textwidth]{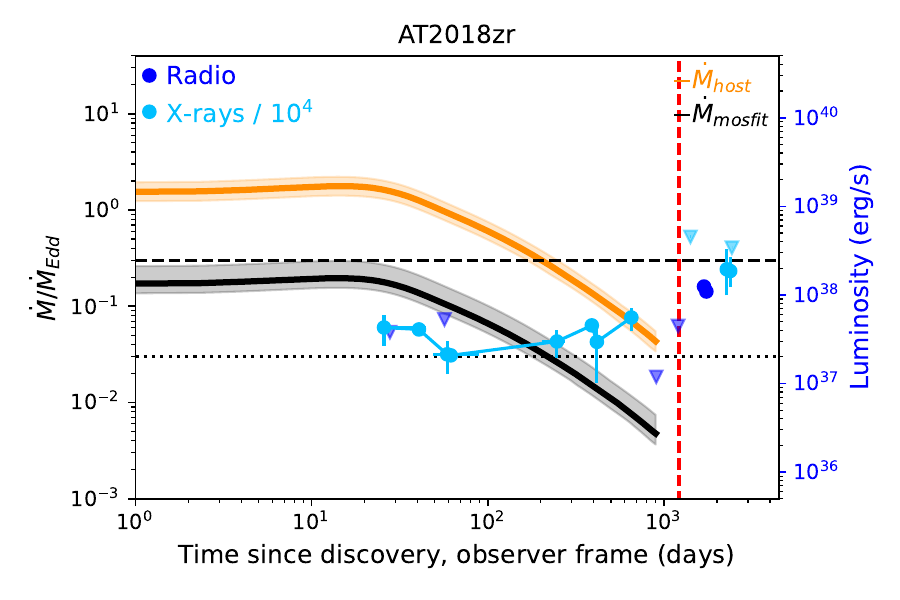}
    
    \vspace{-0.1in}
    
    \includegraphics[width=.48\textwidth]{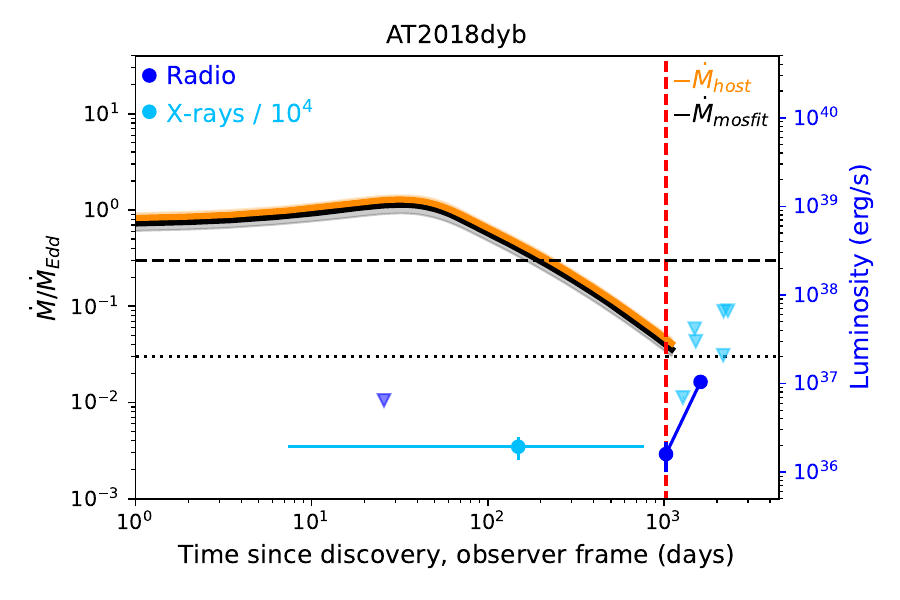}\includegraphics[width=.48\textwidth]{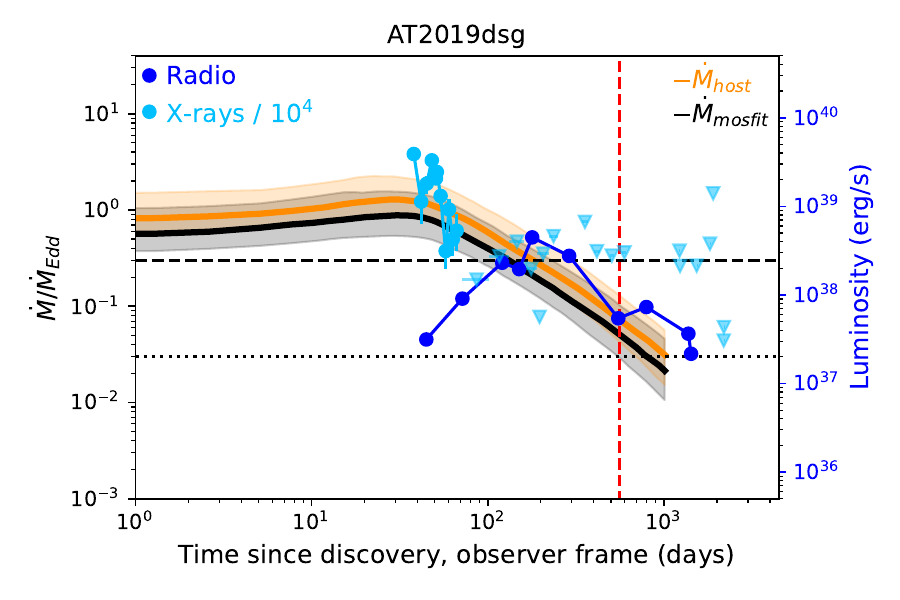}
    
    \vspace{-0.1in}
    
    \includegraphics[width=.48\textwidth]{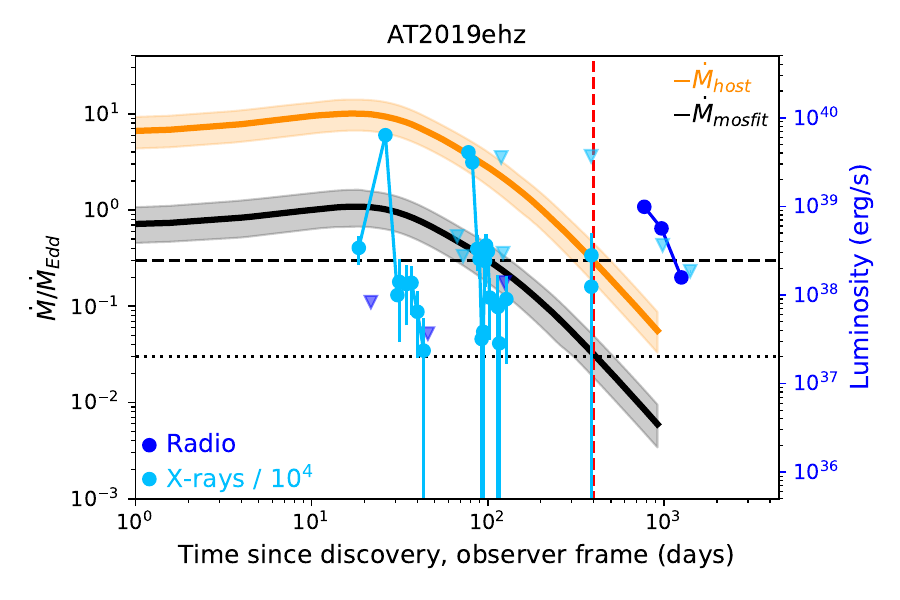}\includegraphics[width=.48\textwidth]{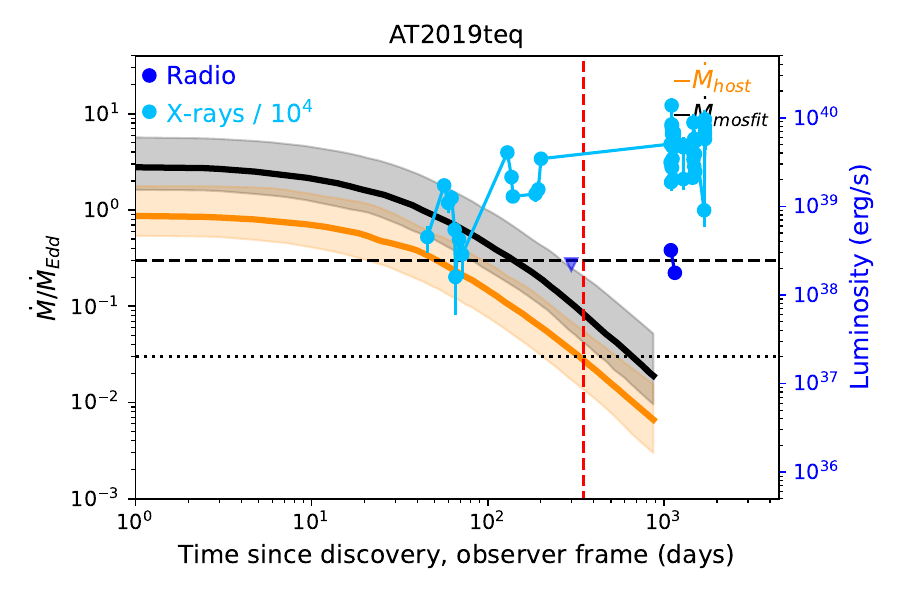}
    
    \vspace{-0.1in}
    \includegraphics[width=.48\textwidth]{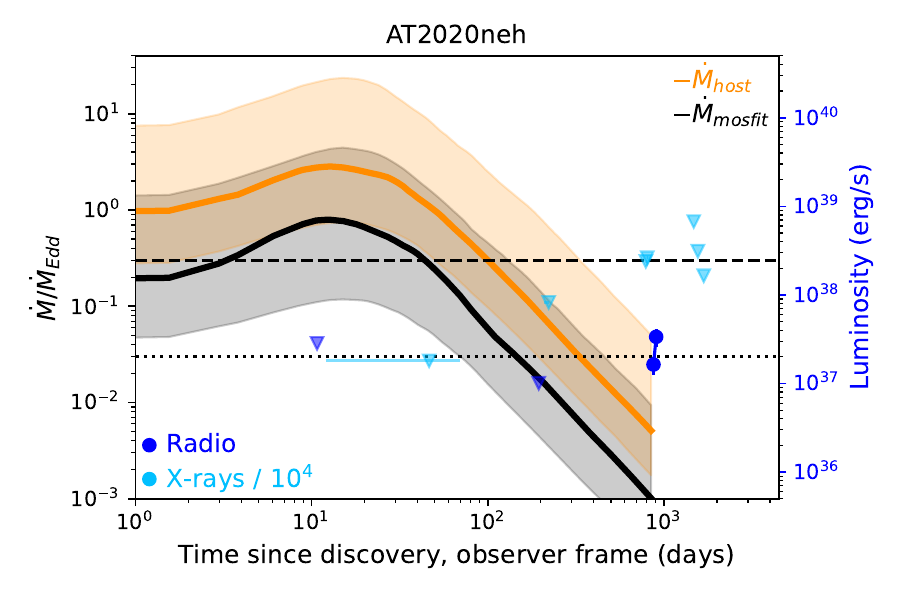}

    \vspace{-0.1in}
    \caption{\label{fig:mdot2}Same as Figure \ref{fig:mdot} for the remaining TDEs in our sample with a delayed radio emission episode and converged MOSFiT models. For six of these TDEs, \yvette\ provide a rough estimate of the outflow launch time (dashed vertical red lines). For AT2019dsg, the dashed line indicates the possible launch time of the second radio flare. For PS16dtm, the open circle denotes a marginal possible detection discovered in our reanalysis of the data (Section \ref{sec:rdata}).  
    \yvette\ do not estimate a launch time for AT2020neh due to its faintness.} 
\end{figure}

\begin{figure}
\centering
    \includegraphics[width=.48\textwidth]{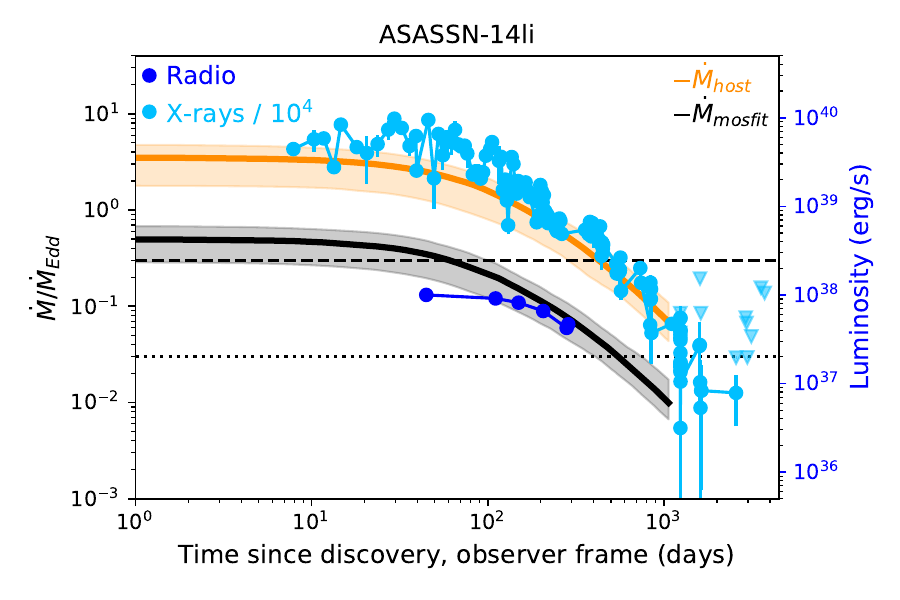}\includegraphics[width=.48\textwidth]{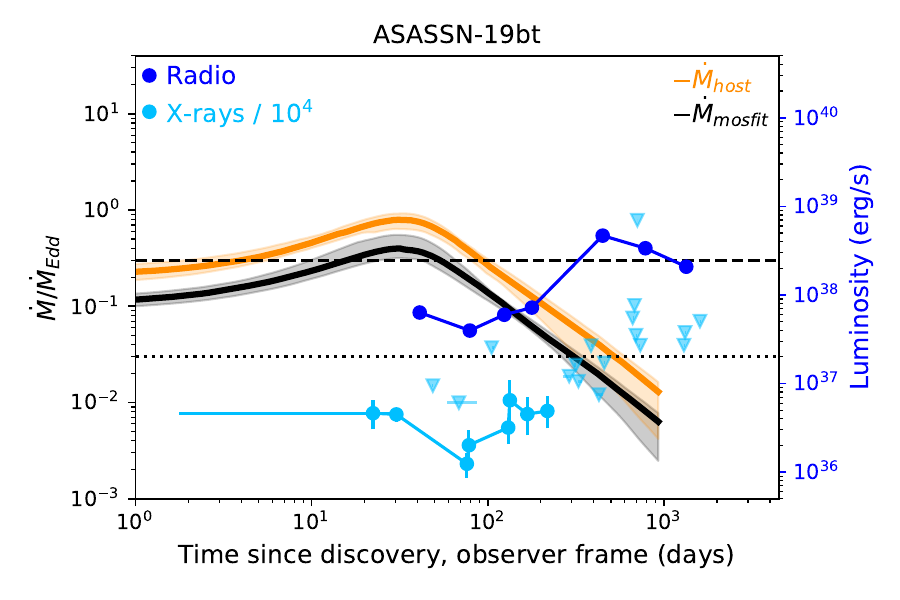}

    \includegraphics[width=.48\textwidth]{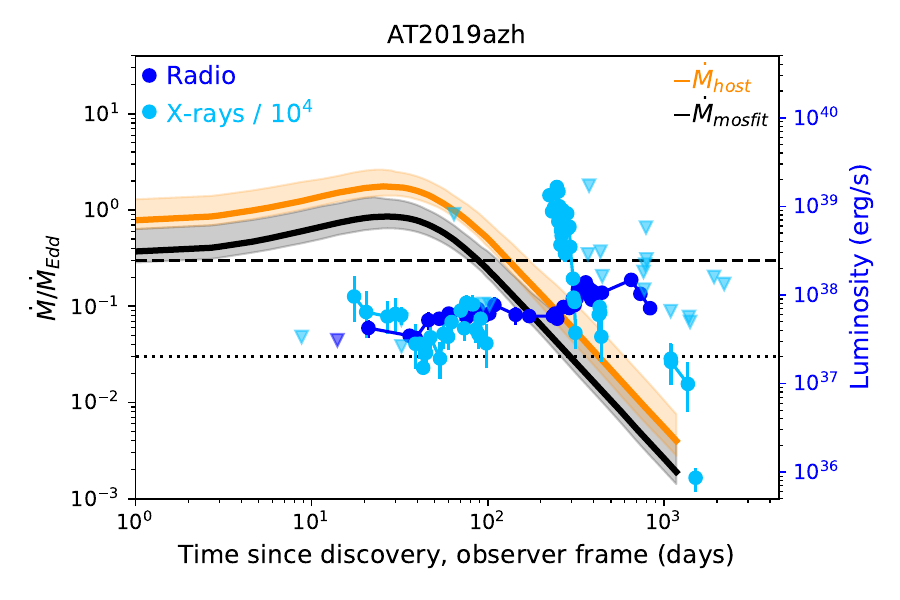}\includegraphics[width=.48\textwidth]{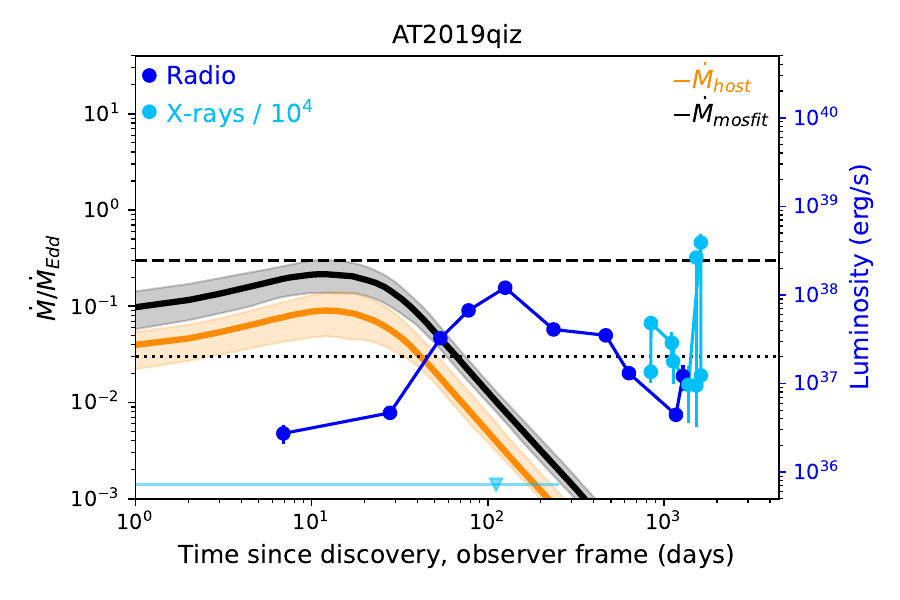}    \includegraphics[width=.48\textwidth]{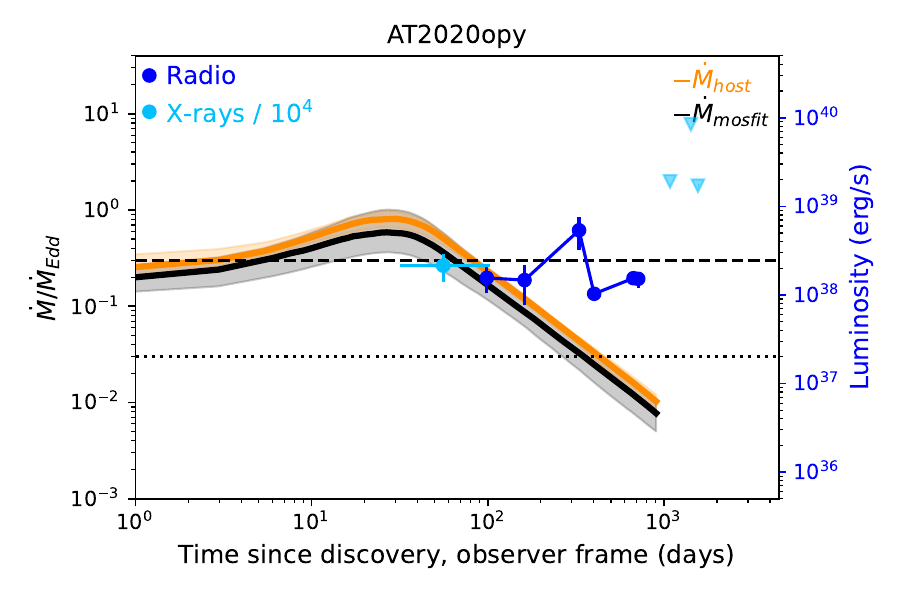}
    \includegraphics[width=.48\textwidth]{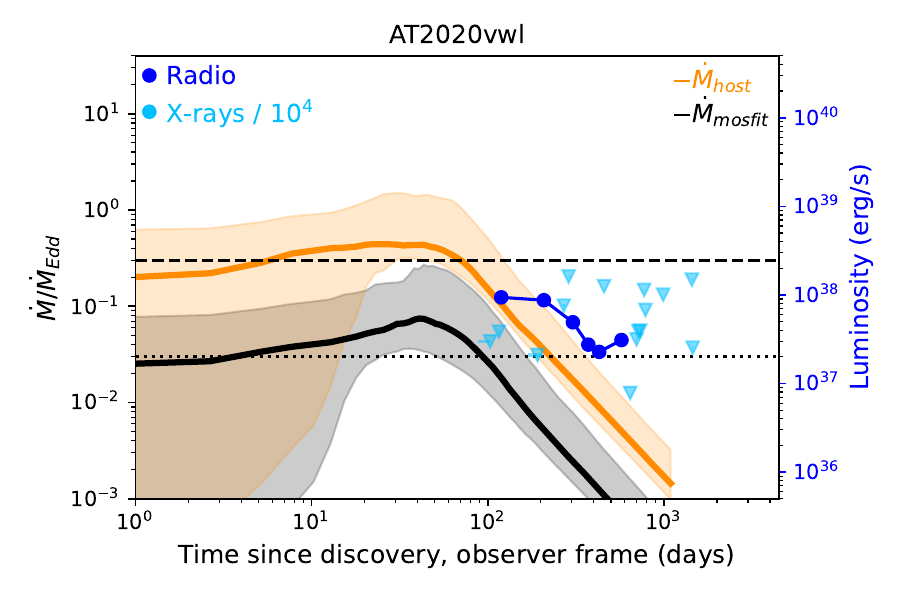}
    
    \caption{\label{fig:mdot3}Same as Figures \ref{fig:mdot} and \ref{fig:mdot2} for our remaining TDEs previously found to exhibit prompt radio emission (within 6 months of discovery). Unlike the TDEs in Figures \ref{fig:mdot} and \ref{fig:mdot2}, several of these TDEs require outflow launch dates corresponding to the initial rise of the optical light curve rather than to a period of super-Eddington accretion (e.g., \citealt{gam+23,caw+24}).  
    This may suggest a different mechanism is required to power early radio outflows in TDEs.} 
\end{figure}

\subsection{MOSFiT modeling: accretion rate}\label{sec:mosfit2}

An additional derived output of the MOSFiT code is a plot of how the accretion rate onto the SMBH, $\dot{M}$, varies with time. To further explore how the radio behavior of TDEs may correlate with the SMBH accretion rate, 
we use MOSFiT to extract the temporal evolution of $\dot{M}$ in our TDEs and compare their observed radio and X-ray evolution to the timing of potential state transitions in the accretion flow (i.e., the times when the accretion disk may switch in or out of a state capable of powering outflows). We assume here that outflows are most likely for super-Eddington accretion ($\dot{M} \gtrsim0.3 \, \dot{M}_{\rm Edd}$ and highly sub-Eddington accretion $\dot{M}\lesssim0.03 \, \dot{M}_{\rm Edd}$). 

In Figures \ref{fig:mdot}-\ref{fig:mdot3} we show the $\dot{M}$ curves together with the radio and X-ray light curves for all TDEs from our sample with unambiguously transient radio emission and converged MOSFiT models\footnote{i.e., all TDEs with radio type ``Prompt'' or ``Delayed'' in Table \ref{tab:gold} except for AT2018fyk, whose optical emission cannot be well-fit by MOSFiT. Further late-time radio monitoring of the TDEs with radio types Host or None is ongoing and will be presented in future work.}. 
Specifically, Figure \ref{fig:mdot} shows TDEs for which we have a reasonably well-constrained estimate of the outflow launch time\footnote{ASASSN-14ae, AT2018hco, AT2018hyz, and AT2019eve have multiple radio SEDs at late times, so it is possible to use the evolution of the derived physical size of the radio-emitting region to obtain a rough estimate of the time at which their delayed outflows were launched (\citealt{cba+22} and \yvette). Similarly, \cite{hsf+21} model the multi-frequency radio light curves of iPTF16fnl to constrain the likely radio outflow launch time to $\sim56-153$ days post-discovery. Multi-wavelength observations of ASASSN-15oi suggest that an outflow launched at the time of the soft X-ray peak may provide a good explanation for its second radio flare \citep{hca21, ham+24}.}, Figure \ref{fig:mdot2} shows TDEs with delayed radio emission where we have at most an approximate constraint on the outflow launch time from \yvette\footnote{plus AT2019dsg, for which we show the approximate onset time of its secondary delayed peak from \yvette, and AT2020neh, which is too faint for an accurate launch estimate.}, and Figure \ref{fig:mdot3} shows the remaining TDEs with prompt radio emission. 
 We normalize $\dot{M}$ to $\dot{M}_{\rm Edd}$, where $\dot{M}_{\rm Edd}$ is the accretion rate required to power emission equal to the Eddington luminosity of the SMBH assuming a radiative efficiency $\epsilon$; $L_{\rm Edd} = \epsilon\dot{M}_{\rm Edd}c^2$. We use the radiative efficiency from MOSFiT for each TDE and our two independent estimates of the SMBH mass from Section \ref{sec:bhmass}.  
 We find that most TDEs in our sample achieve a peak $\dot{M}$ that approaches or mildly exceeds $\dot{M}_{\rm Edd}$. The peak radio luminosity of a TDE does not appear to correlate with the maximum $\dot{M}/\dot{M}_{\rm Edd}$ achieved onto its SMBH. 

We next investigate if the onset time of the radio emission in our TDEs correlates with any specific $\dot{M} / \dot{M_{\rm Edd}}$ ratio, beginning with the TDEs in Figure \ref{fig:mdot} where we have the best constraints.  We find that three of these TDEs (AT2018hco, ASASSN-15oi, and iPTF16fnl) have outflows that were plausibly launched when the accretion rate was a few percent of $\dot{M}_{\rm Edd}$, although for the latter two objects we are limited by the systematic uncertainties on the SMBH masses. ASASSN-15oi's and iPTF16fnl's radio emission may also have been launched earlier, possibly when $\dot{M}$ was still super-Eddington.\footnote{This is particularly likely for ASASSN-15oi, whose multi-wavelength observations suggest that accretion in this TDE may be delayed a few hundred days relative to our MOSFiT prediction \citep{ham+24}.} AT2018hyz and AT2019eve's radio outflows do not appear until several hundred days after the $0.03\,\dot{M}/\dot{M}_{\rm Edd}$ transition, implying that either accretion is delayed relative to the MOSFiT prediction for these two events, or that some other process is responsible for their radio brightenings.\footnote{The launch time from \cite{cba+22} assumes that AT2018hyz's radio outflow is non-relativistic; if its radio emission is instead due to a relativistic off-axis jet, then a much earlier launch time coinciding with the predicted period of super-Eddington accretion is possible \citep{mp23,sbh+24}.} Finally, ASASSN-14ae's very late outflow launch time is difficult to reconcile with either state transition, assuming the outflow launch time and the mass of its SMBH are both correct. This may suggest that either accretion disk formation is significantly delayed in this TDE relative to the prediction from MOSFiT, or some other unrelated process is required to explain its radio rise. 

Although the onset time of radio emission is less clear for the TDEs in Figure \ref{fig:mdot2}, we can roughly approximate the launch time of each outflow based on the radio light curve (\yvette; dashed red lines in Figure \ref{fig:mdot2}).\footnote{We note that our new tentative detection of PS16dtm (Section \ref{sec:rdata}; open circle in Figure \ref{fig:mdot2}) pre-dates the launch time suggested by \yvette\ for this TDE; it is therefore possible that its radio outflow was launched significantly earlier ($\sim1$ year post-discovery).} The systematic uncertainties on the SMBH masses and, in some cases, the sparse temporal sampling of the radio light curves preclude a definitive association of the radio emission with a particular accretion rate for these TDEs. It is nevertheless allowed within the uncertainties that the delayed radio emission in all of these events (including the secondary peak in the light curve of AT2019dsg) is driven by a state change in the accretion flow associated with $\dot{M}$ declining to $\lesssim0.03\,\dot{M}_{\rm Edd}$.\footnote{{PS16dtm and AT2018zr's outflows may also be launched at significantly lower accretion rates, depending on which black hole mass measurement is preferred.}} Again, in most cases, peak accretion onto the SMBH would have to be delayed by several hundred days or more relative to the MOSFiT predictions to associate the outflow launch times with super-Eddington accretion instead.\footnote{PS16dtm and AT2019ehz are the two possible exceptions, if their host galaxy-derived SMBH masses rather than their MOSFiT-derived SMBH masses are assumed to be correct.}

Finally, Figure \ref{fig:mdot3} shows the $\dot{M}$ evolution and radio light curves of the remaining 6 TDEs in our sample for which prompt radio emission was detected within 6 months of discovery. For most of the Prompt radio TDEs, the first epoch of radio follow up unfortunately does not occur until the time of peak optical light or later, as there is typically a delay between transient discovery and the timescale on which most optical TDEs are classified and publicly announced to the community. While direct evidence for pre-peak radio emission in these TDEs is therefore limited,  AT2019qiz's first radio detection (7 days post-discovery, the earliest radio detection of any TDE in our sample) occurs when $\dot{M}$ is sub-Eddington and has not yet peaked, suggesting that the process powering prompt radio emission in this TDE may not be related to a high accretion rate onto the SMBH. Additionally, for several events (ASASSN-19bt, AT2019dsg, AT2020vwl), equipartition analyses of their multi-frequency radio SEDs suggest that the launch times of their radio outflows coincide with the time their optical light curves began to rise -- well before $\dot{M}$ peaks \citep{cab+21,gam+23,caw+24}.

\subsection{X-ray light curves and the Fundamental Plane}\label{sec:fp}

\begin{figure}
\centering
    \includegraphics[width=.8\textwidth]{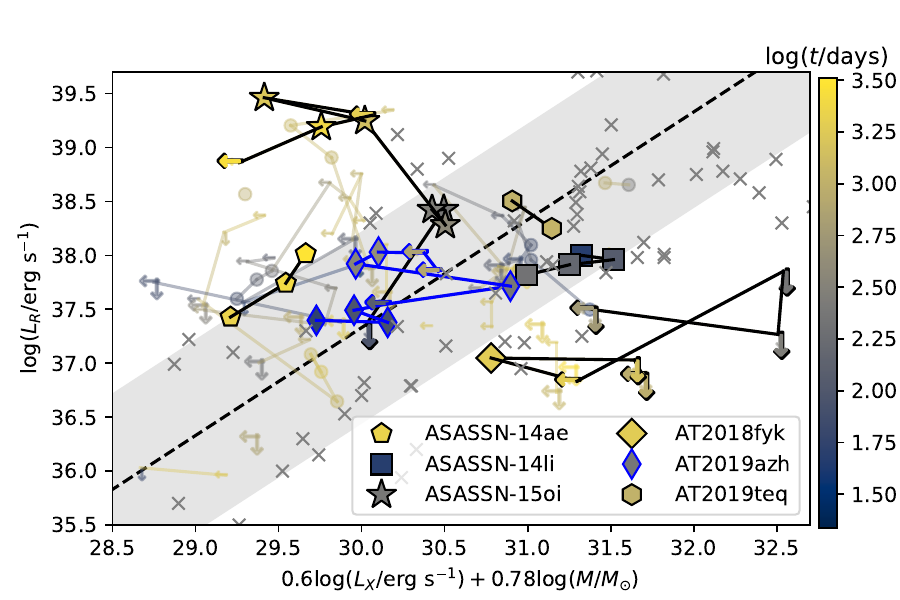}
    \caption{\label{fig:fp}The quasi-simultaneous ($\delta t/t < 0.25$) X-ray and radio luminosities of our TDEs shown relative to the Fundamental Plane relationship (dashed black line with intrinsic scatter given by the shaded gray region) and AGN sample (gray X's) from \cite{m03}. Arrows indicate upper limits and points from the same TDE are connected by a solid line. The color of each point indicates the time in days since discovery (dark blue = earliest, yellow = latest). 
     We highlight six particularly noteworthy TDEs with unique symbols given in the figure legend. We find that X-ray and radio-detected TDEs generally follow the fundamental plane at early times (within 6 months of discovery), as do a subset of the TDEs with radio detections at late times. ASASSN-15oi (stars) is consistent with the plane during its first radio flare but deviates from the plane during its second radio flare, possibly suggesting different physical origins for the radio emission on these two timescales.}
\end{figure}
 
The X-ray properties of our TDEs may also connect to the instantaneous accretion rate onto the SMBH (e.g., \citealt{m22,sm24}). Together with our SMBH mass measurements (Section \ref{sec:bhmass}), we can use our radio and X-ray light curves to track the evolution of individual TDEs in the $L_X$-$L_R$ phase space over thousands of days and compare the properties of our sample to the Fundamental Plane of black hole activity. The Fundamental Plane is a well-established correlation between radio luminosity, X-ray luminosity, and black hole mass that holds across many orders of magnitude for black holes accreting at low rates \citep{m03,fkm04}. If delayed radio emission is produced in TDEs after the accretion rate has dropped to similarly low levels, we may expect these TDEs to obey the same relation at late times.

We show our TDE sample on the Fundamental Plane in Figure \ref{fig:fp}, using the SMBH masses derived from the host galaxy properties\footnote{Using $M_{\rm BH,host}$ rather than $M_{\rm BH,mosfit}$ allows us to include DES14C1kia and AT2018fyk, which do not have converged MOSFiT models.} (Table \ref{tab:bhmass}). The original sample of AGN from \cite{m03} are also shown for comparison. As this relation is calibrated using radio observations collected at 5 GHz, we restrict ourselves to radio data taken at similar frequencies ($\nu < 10$ GHz). We furthermore only show TDEs for which quasi-simultaneous radio and X-ray data are available, i.e., observations for which $\delta t/t < 0.25$. For TDEs with high-cadence observations, if multiple radio/X-ray data point pairs fall within this tolerance, we select the data pairs with the smallest separation in time and frequency space, such that each available data point is used at most once in the figure. {29 of our 31 TDEs have at least one radio/X-ray data point pair satisfying these restrictions and are included in Figure \ref{fig:fp}.} We highlight six TDEs in the figure legend {that illustrate the diverse range of radio and X-ray behaviors seen in our sample.} Among the TDEs with sufficiently good radio and X-ray coverage, we see that all of our prompt radio TDEs and many of our delayed radio TDEs are consistent with the Fundamental Plane {(22 of 29 events)}. Conversely some TDEs move away from the Plane at late times. {The existing data show clear movement away from the plane for two TDEs (ASASSN-15oi and AT2018hyz). Several other TDEs may also exhibit this trend (most suggestively ASASSN-19bt, also potentially PS16dtm and AT2019ehz), but the limitations of the current data preclude a definitive analysis.}

\section{Discussion}\label{sec:disc}

We now attempt to synthesize our multi-wavelength analyses into a physical model, or models, to explain delayed radio emission in TDEs. Any model must be able to explain our key observational findings from Section \ref{sec:mod}, which we briefly summarize here:

\begin{itemize} 
\item TDEs with delayed radio emission are more likely to entirely lack He emission in their optical spectra or to only exhibit it at late times, in comparison to TDEs with prompt radio emission ($p=0.002$).
\item TDEs with delayed radio emission have larger photospheric radii than TDEs with prompt radio emission ($p=0.026$).
\item {Although we are limited by systematics in our SMBH mass measurements, 3-10 of our} TDEs (e.g., AT2018hco, the second radio flare in AT2019dsg) have radio outflow launch times that roughly coincide with MOSFiT's predicted transition to a highly sub-Eddington state $\dot{M}<0.03\,\dot{M}_{\rm Edd}$. However, we find that {3-6} delayed radio outflows are launched up to several hundred days later than the MOSFiT predictions for the timing of this state transition (e.g., in ASASSN-14ae). 
\item Similarly, while the radio launch times of a few TDEs in our sample may be marginally consistent with MOSFiT's predicted super-Eddington phase $\dot{M}>0.3\,\dot{M}_{\rm Edd}$ (ASASSN-15oi's first flare, iPTF16fnl, PS16dtm, AT2019ezh), the radio emission in most of our delayed radio TDEs can only be explained as outflows powered by super-Eddington accretion if the super-Eddington phase is delayed by several hundred days relative to the MOSFiT prediction.
\item The onset of prompt radio emission in some TDEs precedes the period of peak accretion onto the SMBH.
\item {$22$ of our} TDEs, including all TDEs in our sample with prompt radio emission and some TDEs with delayed radio emission, are broadly consistent with the Fundamental Plane of black hole activity.
\item The two TDEs that show the most dramatic late-time radio brightenings (AT2018hyz and ASASSN-15oi's second flare) are both \emph{inconsistent} with the Fundamental Plane at late times.
\end{itemize}

Broadly speaking, there are two classes of models that have been proposed to explain radio emission in TDEs: models in which the radio emission is directly powered by accretion-driven outflows (e.g.,~\citealt{sq09,gm11,dgnr12,abg+16,vas+16,m22}), and those where it originates from some other aspect of the disruption process unrelated to SMBH accretion. The latter category includes emission from shocks between the unbound portion of the stellar debris stream and the ambient environment \citep{kps+16}, emission from outflows launched by stream-stream collisions during the circularization of the bound stellar debris \citep{lb20}, and sudden brightenings of a shock that encounters density variations in the circumnuclear environment \citep{zsml24}. In the first category, we might expect correlations between the radio behavior and other probes of the accretion rate (such as the optical or X-ray emission), while in the second category we would not expect such correlations. As we explore in this section, the existence of the various multi-wavelength connections uncovered in our analysis thus suggest that the late-time radio properties of our sample are driven by accretion onto the SMBH, rather than by environmental interactions.

\subsection{Overview of accretion-driven jets and outflows in TDEs}

As discussed in \citet{gm11}, there are two phases after a TDE where an accretion-driven relativistic jet or other radio-producing outflow may be formed.  Firstly, if the process of debris circularization and disk formation is sufficiently rapid, then a jet may accompany the early super-Eddington phase ($\dot{M} \gg \dot{M}_{\rm Edd})$, in analogy with gamma-ray bursts. Such an early, super-Eddington jetted phase is likely responsible for highly energetic jetted TDEs such as Swift J1644+57 \citep{dgnr12}.  This early jet phase ends once the disk becomes geometrically thin at $\dot{M} \lesssim 0.3\,\dot{M}_{\rm Edd}$ and can no longer retain its magnetic flux (e.g., \citealt{msn+06,tmg+14}).  This transition typically occurs hundreds of days post-disruption, consistent with the observed shut-off time of the X-rays in jetted TDEs \citep{zbm+13,ckh+12,eta+24}. Super-Eddington accretion has also been proposed to power non-relativistic winds in TDEs that produce low-luminosity radio emission at early times (e.g., \citealt{sq09,abg+16}). The presence of a wind may explain why only powerful super-Eddington jets have been seen at early times: if the star's orbital plane is initially misaligned with the black hole spin axis, the jet will precess and less powerful jets may only be able to break out of the wind after the disk undergoes sufficient Lense-Thirring precession to come into alignment with the spin \citep{tm23,lmm24}. This will then delay any radio emission from less powerful super-Eddington jets. Radio emission from powerful relativistic jets may also appear delayed if the jets are viewed off-axis, as relativistic beaming will suppress the emission at early times. {For nearby TDEs, off-axis jets may be directly resolved with Very Long Baseline Interferometry (VLBI) observations, providing a clear method for distinguishing between jetted and non-jetted emission (e.g., \citealt{mpe+18}).}

Alternatively, a jet may form at very late times, once the accretion rate decreases to $\dot{M} \lesssim 0.03\,\dot{M}_{\rm Edd}$ \citep{gm11}. When the accretion rate becomes this low, the accretion flow is predicted to transition back to radiatively inefficient and geometrically thick, which can again power outflows (e.g., \citealt{ny95,bb99}).  Such behavior is supported by detections of radio jets from X-ray binaries (XRBs) in the low-hard state and bright radio emission consistent with compact jets from AGN at low Eddington ratios (e.g., \citealt{fct+99,fkm04,ssl07,ccb+13}). The rapid X-ray and optical spectral and photometric variability observed in changing-look AGN \citep{mgm03,rt23} also supports the existence of this state change, as this variability is consistent with the sudden appearance or disappearance of a jet and changing-look AGN are predominantly observed to have Eddington ratios of a few percent, right at this expected transition. (This is much lower than the typical Eddington ratios seen in the general AGN population; \citealt{mga+19,gpa+22,gzg+24}.) A delayed radio jet was recently discovered in a changing-look AGN $\sim6$ years after its original optical outburst, providing direct evidence that these events can power outflows \citep{mls+25}. In TDEs, this late jet phase is expected to occur on timescales of years to decades.  

We expect our TDEs to adhere to the Fundamental Plane of black hole activity (Section \ref{sec:fp}) during periods where their SMBHs are accreting at a low fraction of the Eddington rate \citep{fkm04,jsg+20}. Conversely, if the accretion rate is still high during some observed delayed radio brightenings, then we might expect the radio and X-ray emission to deviate from the Fundamental Plane for these TDEs. We may also expect to see a given TDE evolve along the Plane or towards/away from it as $\dot{M}$ changes and the accretion flow possibly undergoes state changes.  The radio and X-ray emission of some TDEs also may not follow the Fundamental Plane at any epoch if the radio emission is not directly driven by accretion onto the SMBH (e.g., if prompt emission is produced instead by the shock between the unbound portion of the stellar debris stream and the ISM; \citealt{kps+16}, and/or if late-time emission is due to a previously-launched outflow encountering a change in the ambient environment; \citealt{zsml24}). In the case of radio emission dominated by a precessing jet whose escape is delayed due to an initial misalignment between the disk and the SMBH spin, we also do not expect any clear correlation between the simultaneous radio and X-ray emission, as the radio onset time is determined by the alignment timescale rather than the accretion rate in these models \citep{tm23,lmm24}.

\subsection{Origin(s) of the delayed radio emission}\label{sec:delay}

Our delayed radio TDEs exhibit radio emission on a wide range of timescales (hundreds to thousands of days post-discovery) and with a wide range of luminosities, suggesting that a single mechanism may not be sufficient to explain the entire sample. In this section, we discuss the various types of accretion-driven outflows that may contribute to the delayed radio emission. As described in Section \ref{sec:mod}, we constrain the accretion rate as a function of time for the TDEs in our sample via two methods: modeling the optical/UV light curves using MOSFiT, and comparing their X-ray and radio properties directly to the Fundamental Plane of black hole activity. 

\subsubsection{(Delayed) super-Eddington accretion}
We first consider scenarios in which the delayed radio emission observed in our TDE sample is powered by non-relativistic outflows launched during a phase of super-Eddington accretion. While this is unlikely to explain TDEs with radio emission delayed by thousands of days (e.g., ASASSN-14ae)\footnote{It is possible to delay peak accretion onto the SMBH by this much under certain assumptions (for example in the models for low-mass SMBHs explored by \citealt{gr15}, where the debris stream may only self-intersect after dozens of pericenter passes and debris circularization can therefore be delayed by years). However, delaying the stream-stream collisions and debris circularization phases to this extent also spreads out the resulting mass fallback such that the peak accretion rate is likely no longer super-Eddington, and the optical evolution is likely much slower than observed in ASASSN-14ae.}, it may be a plausible explanation for TDEs where a radio outflow is launched hundreds days after discovery.

Many of our delayed radio TDEs have larger optical photospheric radii at early times and delayed or absent He lines. {As noted in Section \ref{sec:opt},} previous work by \cite{nlr+22} found that TDE-H may arise from either less complete stellar disruptions (i.e., those with a smaller impact parameter, $b$) or from those in which an accretion disk does not form promptly. Our observed correlation between delayed radio emission and a {TDE-H classification} could thus suggest that less complete stellar disruptions are also more likely to produce delayed radio emission, but we disfavor this interpretation as we find no statistically-significant difference in the $b$ values of our Prompt and Delayed TDE sub-samples {(Table \ref{tab:ad})}. Instead, the fact that our delayed radio subsample both lacks prompt He emission and has larger optical photospheres favors the delayed accretion hypothesis. \cite{nlr+22} also found that their TDE-H exhibit larger photospheric radii than other types of TDEs, consistent with previous studies \citep{hhs+20,vgh+21,clm+22}. \cite{nlr+22} suggest that this could imply the early optical emission in TDE-H could be powered by collision-induced outflows rather than promptly formed accretion disks, such that accretion onto the SMHB is delayed relative to peak optical light. This would also naturally delay the onset of radio emission produced by accretion-powered outflows.

If the optical and UV emission around peak light is produced by energy dissipated at the debris stream self-intersection point, as suggested by e.g., \cite{psk+15,jgl16}, then larger photospheric radii would correspond to TDEs where the self-intersection point is located further away from the SMBH. \cite{nlr+22} suggest that this could occur in TDEs around less massive SMBHs. In such TDEs, it seems plausible that it would take correspondingly longer for the debris to reach the event horizon, resulting in delayed peak accretion onto the SMBH. Our weak correlation between delayed radio emission and larger $R_{\rm ph,0}$ values (and our even more tentative correlation between delayed radio emission and smaller $M_{\rm BH,host}$ values) could then suggest that at least some delayed radio outflows in TDEs are accretion-driven, and specifically are powered by a phase of high (super-Eddington) accretion delayed several hundred days relative to peak optical light.\footnote{While our sample also exhibits the inverse correlation between $R_{\rm ph,0}$ and SMBH mass seen by \cite{nlr+22} (Figure \ref{fig:Rph0M*}, right), it is interesting to note that this correlation is not seen in the recent flux-limited TDE sample studied by \cite{yrg+23}, who instead find a weak positive correlation between the blackbody radius at peak optical light and the SMBH mass.} 

Alternately, the stellar debris may circularize rapidly without immediately accreting. Instead, the circularized debris may initially form a hot quasi-spherical pressure-supported envelope \citep{ss24,plm+24}, delaying rapid accretion onto the SMBH until the envelope cools and contracts to form a disk \citep{lu97,cb14,m22}. If the pressure-supported envelope model is correct, the timing of the super-Eddington accretion phase may be delayed by several hundred days relative to the MOSFiT models presented in Section \ref{sec:mosfit}, suggesting that it may still be possible for super-Eddington accretion to power the delayed radio emission observed in some TDEs even if the radio onset time occurs after MOSFiT's nominal $\dot{M}\sim0.3 \; \dot{M_{\rm Edd}}$ transition point rather than prior to it \citep{m22,sm24}. In this case, the peak of the X-ray light curve will also be delayed relative to peak optical light and may provide a better estimate of the peak accretion time \citep{m22}. This is a likely explanation for the delayed X-rays seen in ASASSN-15oi (Figure \ref{fig:mdot} and \citealt{ham+24}). In scenarios where the early TDE UV/optical emission instead originates from a quasi-spherical envelope \citep{m22}, the initial size of the envelope (and hence its photosphere radius) increases with the SMBH mass; however, the cooling timescale over which the envelope contracts is also longer for lower SMBH mass, such that in the simplest scenario of passive cooling, the photosphere radius at a fixed time after the disruption is roughly independent of the SMBH mass. Given the low significance of our possible correlation between SMBH mass and delayed radio emission, this scenario is also consistent with our results.

\subsubsection{Relativistic jets}

Super-Eddington accretion can also power radio-luminous relativistic jets \citep{dgnr12}. The two TDEs that show the most dramatic late-time radio brightenings (AT2018hyz and ASASSN-15oi) are both \emph{inconsistent} with the Fundamental Plane at late times. This is clearest for ASASSN-15oi (stars in Figure \ref{fig:fp}), which shows two distinct radio emission episodes (Figure \ref{fig:mdot}). ASASSN-15oi's emission is consistent with the Fundamental Plane prior to and during the first radio flare, but moves significantly further away from the plane during the second flare, as the X-rays fade at the same time that the radio dramatically brightens. This suggests two different physical origins for the two radio flares, as proposed by \cite{ham+24}. Several other TDEs in our sample possibly show a similar evolution away from the Fundamental Plane at late times (AT2019ehz, PS16dtm, and ASASSN-19bt), which could be confirmed by continued late-time radio monitoring accompanied by deeper X-ray observations. 

The rapid radio brightening seen in these events may be powered by a jet (or wind) launched during a period of high accretion onto their SMBHs, when they are not expected to be on the Fundamental Plane. This is broadly consistent with the MOSFiT accretion rate predictions at the outflow launch times for ASASSN-15oi, PS16dtm, and AT2019ehz shown in Figures \ref{fig:mdot} and \ref{fig:mdot2}, although the systematic uncertainties on the SMBH masses for all three TDEs are large. The launch time for AT2018hyz shown in Figure \ref{fig:mdot} is taken from the non-relativistic outflow modeling by \cite{cba+22} and does not coincide with MOSFiT's prediction for super-Eddington accretion. This may either suggest that MOSFiT places the time of peak accretion too early, or that AT2018hyz's radio emission is powered by an off-axis jet launched much earlier than the \cite{cba+22} prediction but not visible until late times because of relativistic beaming \citep{mp23} or delayed disk-black hole spin alignment \citep{tm23,lmm24}. Further radio monitoring of AT2018hyz, including VLBI observations, may distinguish among these possibilities. 

We also note that for TDEs, which exhibit rapidly-evolving accretion rates rather than the steady-state accretion typically seen in AGN, simultaneous radio and X-ray observations may not always be probing the same accretion rate if the radio emission is significantly spatially separated from the X-rays. In AGN on the Fundamental Plane, the radio emission is typically dominated by flat-spectrum emission from the jet core \citep{fkm04} and thus reflects the current (or recent) accretion rate, as does the X-ray emission (which is presumed to originate close to the SMBH in both AGN and TDEs). In contrast, if the radio emission in our TDEs is dominated instead by optically-thin synchrotron emission from the forward shock between the ISM and the leading edge of an expanding jet (as first proposed for Swift J1644+57; e.g., \citealt{zbs+11}), then it may instead reflect the accretion rate at the time the jet was launched, which could be significantly different from the current accretion rate. This could explain why ASASSN-15oi and AT2018hyz deviate from the Fundamental Plane even at very late times. A similar argument has previously been suggested to explain why AGN don't always follow the Fundamental Plane at low SMBH masses similar to those of TDE hosts \citep{gng+22}.

\subsubsection{Highly sub-Eddington accretion}\label{sec:super}

Some delayed radio emission in TDEs may also be powered by highly sub-Eddington accretion. To assess this possibility, we must accurately determine the timing of the $\dot{M} \; \sim0.03\,\dot{M}_{\rm Edd}$ state transition, when the disk is expected to transition from geometrically thin back to geometrically thick. In our Case 1 prompt accretion scenario, we can straightforwardly compute this timescale using our $\dot{M}$ curves from MOSFiT (Section \ref{sec:mosfit2}). In Case 2 delayed accretion scenarios, the situation is less clear cut. If the cooling envelope model developed by \cite{m22} is correct, then assuming that the envelope contracts to a disk faster than the mass fall-back reaches $\dot{M} \; \sim0.03\,\dot{M}_{\rm Edd}$, and neglecting any spreading disk phase, the timescale estimates for this state transition that we obtain with MOSFiT while neglecting the envelope phase may still be accurate.\footnote{We note that even if the MOSFiT prediction for the timing of the $\dot{M} \; \sim0.03\,\dot{M}_{\rm Edd}$ transition is underestimated, parameters such as $R_{\rm ph,0}$ that are primarily determined by the photometry around peak optical light are still reliable.} We find that some TDEs in our sample (e.g., AT2018hco) do indeed exhibit delayed radio emission consistent with an outflow launched at approximately the same time that MOSFiT predicts this state transition will occur (Figure \ref{fig:mdot}).

However, the existence of UV plateaus in many TDEs that remain well-fit by a thin disk model for years post-disruption (e.g., \citealt{vsm+19,mb20,mnig24,mvn+24}) may suggest that in reality the $\dot{M} \; \sim0.03\,\dot{M}_{\rm Edd}$ transition is also delayed relative to the MOSFiT prediction, at least for some TDEs.\footnote{This is not always the case; \cite{mnig24} apply the thin disk model FitTeD to AT2019dsg and find that the onset of its second radio flare approximately coincides with the calculated $\dot{M} \; \sim0.03\,\dot{M}_{\rm Edd}$ state transition, just as MOSFiT does (Figure \ref{fig:mdot2}).} Furthermore, in contrast to the hard X-ray spectra typical of AGN accreting at highly sub-Eddington rates, the X-ray emission from TDEs on timescales of years is often soft, consistent with thermal thin-disk emission extending to late times (e.g., \citealt{agr17,gca17,wjs+20,jsg+20,ggy+23} -- but see also \citealt{khs+04,skaj20} for discussion of X-ray selected TDEs that do show spectral hardening on timescales of years). We therefore must consider the full multiwavelength properties of each TDE in our sample to assess whether MOSFiT can provide an accurate estimate of its state transition timing, or whether other accretion probes (such as the X-ray emission) should be preferred. 

The strongest evidence that some of our delayed radio TDEs may have entered the highly sub-Eddington phase comes from comparing their radio and X-ray emission. Many of our delayed radio TDEs (e.g., ASASSN-14ae, AT2018fyk, and AT2019teq) align with the Fundamental Plane at late times (Figure \ref{fig:fp}). This may indicate that these TDEs have indeed transitioned to an analog of the low-hard state seen in XRBs at late times and, like the AGN and XRBs used to calibrate the original Fundamental Plane relationship, are accreting at a small fraction of the Eddington rate at the time of observation. Indeed, AT2018fyk's and AT2019teq's X-ray spectra show significant hardening at late times, consistent with this picture \citep{Wevers19, wpv+21,yg22, ggy+23,pcg+24}. This is also consistent with the radio non-detections of AT2018fyk at very early times when the X-rays were significantly brighter and softer, as outflows are not expected prior to this state transition when the accretion rate is higher but likely not super-Eddington given the high SMBH mass {(\citealt{fct+99,fg14};} \citealt{Wevers19}). ASASSN-14ae is the only TDE in our sample to show a correlated increase in both radio and X-ray emission at late times (Figure \ref{fig:mdot}), but its late-time X-ray emission remains too faint for a detailed spectral analysis \citep{jsg+20}. ASASSN-14ae is also the oldest TDE in our sample and has the longest delay between peak optical light and first radio detection. As illustrated by Figure \ref{fig:mdot}, in order to explain ASASSN-14ae's radio emission as an outflow launched by a state transition in the accretion flow at $\dot{M}\sim0.03\,\dot{M}_{\rm Edd}$, accretion onto the SMBH must be significantly delayed relative to the $\dot{M}$ curve computed by MOSFiT (by $\sim2000$ days). Thus, further monitoring of ASASSN-14ae's radio and X-ray emission is necessary to ascertain the nature of the late radio rise in this TDE.

One caveat to this picture is that while some models predict a correlation between the between the SMBH mass and the timescale required to reach the low-hard state transition (typically hundreds of days; e.g., \citealt{dgnr12,m22}), we observe no clear correlation in our data between the SMBH mass and the onset or peak timescales of the radio emission, or between the SMBH mass and the peak radio luminosity. However, as noted in Section \ref{sec:mosfit} and \yvette, the exact onset time of radio emission is poorly constrained for some events. In particular, seven of our TDEs have no radio observations at $t<180$ d and others are only sparsely observed on these timescales, meaning that their prompt radio emission could have been missed entirely. If these TDEs were misclassified in our analysis, then the simple statistical tests we conduct here may not accurately reflect the underlying population. We therefore do not rule out these models immediately based on the lack of an obvious trend between SMBH mass and radio onset time, but instead choose to defer a more in-depth analysis to when a larger sample of TDEs with well-sampled radio light curves at both early and late times is available.

\subsection{Origin of the prompt radio emission}

Finally, we briefly consider the origin of prompt radio emission in our TDEs. While various processes can delay accretion onto the SMBH relative to the $\dot{M}$ curves predicted by MOSFiT (e.g., \citealt{m22}), it is more difficult to devise a mechanism by which the SMBH accretion could peak \emph{earlier} than the MOSFiT prediction. This suggests that some other mechanism occurring earlier in the disruption process may be required to explain radio emission in some TDEs at early times -- perhaps collision-induced outflows launched during the initial circularization of the stellar debris stream \citep{dec13,gr15,jgl16,stg+16,lb20}, or the interaction of the unbound portion of the stellar debris stream with the circumnuclear medium \citep{kps+16,yspk19}. {Both of these models can explain the range of radio luminosities observed in TDEs at early times.} However, in the latter model we would not expect any correlation with the Fundamental Plane, in contrast to our observations (Figure \ref{fig:fp}).  

The observed adherence of our prompt radio TDEs to the Fundamental Plane may instead indicate that a small amount of stellar material does manage to accrete onto the SMBH even at early times. However, the X-ray emission of most Prompt radio TDEs, including the canonical event ASASSN-14li (squares in Figure \ref{fig:fp}), is much softer than that of the AGN that define the Fundamental Plane, arguing against a standard XRB-like low-hard state for this early phase \citep{pv18,ggy+23}. Our other highlighted Prompt radio TDE, AT2019azh, has one of the most extensive radio and X-ray datasets of any TDE; the data included in Figure \ref{fig:fp} span $\sim40-800$ days post-discovery (blue-edged diamonds). Interestingly, it remains roughly consistent with the Fundamental Plane for this entire time, despite late-time changes in its X-ray hardness that suggest possible XRB-like state changes in its accretion flow \citep{shf+22}. The proximity of the Prompt radio TDEs to the Fundamental Plane could conceivably be a coincidence given the small sample size, as noted by \cite{pv18}.  
Indeed, the X-ray non-detections of AT2020vwl, AT2019qiz (at early times), and ASASSN-19bt (at late times) are potentially consistent with deviations from the Fundamental Plane, but deeper observations would be required to confirm this. As the number of TDEs in the published literature with sufficiently sensitive high-cadence radio and X-ray detections at early times remains extremely small, further exploration of these possibilities will require a larger sample of promptly-observed radio TDEs and is beyond the scope of this work.

\section{Conclusions}\label{sec:conc}

We present an in-depth multi-wavelength analysis of a sample of 31 optical TDEs exhibiting radio emission on a variety of timescales, with the goal of elucidating the process(es) responsible for delayed radio emission in some events. We find that the multi-wavelength properties of our sample are as diverse as their radio properties, but nevertheless accretion-powered outflows remain an attractive option to explain delayed low-luminosity radio emission in many TDEs. Our primary conclusions are the following:

\begin{itemize}

\item We use the TDE modeling code MOSFiT to predict the accretion rate onto the SMBH as a function of time for each TDE and estimate the timing of likely state transitions in the accretion flow. For {at least 3, and up to 10} TDEs (e.g., AT2018hco, the second radio flare in AT2019dsg), the transition to a highly sub-Eddington state $\dot{M}<0.03\,\dot{M}_{\rm Edd}$ is roughly coincident with the launch time inferred for the delayed radio outflow, possibly suggesting that their radio emission is related to this state transition. 

\item {A further 3-6} TDEs may have radio outflows launched coincident with the $\dot{M}\sim0.03\,\dot{M}_{\rm Edd}$ state transition if this transition is \emph{delayed} by several hundred days relative to the MOSFiT prediction. This is also supported by the X-ray behavior of these TDEs, which are consistent with the Fundamental Plane of black hole activity at late times. For ASASSN-14ae, this transition may happen as late as $\sim2000$ days post-discovery.

\item Other TDEs (ASASSN-15oi, AT2018hyz) are much more radio-bright/X-ray dim than is typical for black holes accreting at a few percent of the Eddington rate, as seen by their increasing distance off the BH fundamental plane with time (Figure \ref{fig:fp}). We suggest that this sub-group of delayed radio TDEs is powered instead by jets or outflows launched during a period of super-Eddington accretion. For these TDEs, the lack of bright radio emission at early times then implies either that the radio emission is powered by a promptly-launched off-axis jet initially suppressed by relativistic beaming, or that super-Eddington accretion is delayed by several hundred days relative to peak optical light.

\item TDEs with delayed radio emission may be identifiable at early times by their larger optical photospheric radii and lack of He emission at peak. If this indicates slower debris circularization and delayed-onset accretion as previously suggested by \citep{nlr+22}, the super-Eddington accretion phase would also be later than MOSFiT's prediction and some delayed radio TDEs could be powered by non-relativistic outflows launched during this phase. {A super-Eddington phase delayed by up to several hundred days relative to MOSFiT's prediction could explain the radio emission seen in 11/12 of our delayed radio TDEs plus the second flare in AT2019dsg; the exception is ASASSN-14ae.} If this correlation between TDE optical and radio properties is confirmed in a larger sample of TDEs, it may provide an important tool to predict which TDEs are likely to brighten in the radio at late times, enabling more efficient use of limited follow up resources.

\item Taken together, \emph{our analyses suggest that in TDEs the accretion rate commonly peaks several hundred days after peak optical light and then declines}, driving radio-bright outflows in both the super-Eddington and highly sub-Eddington regimes. This may support delayed circularization models for the optical emission or the recent cooling envelope model proposed by \cite{m22}, in which the debris circularizes promptly but does not actually accrete until it has sufficiently cooled. 
    
\item Finally, while a definitive assessment is precluded by the scarcity of available data, the onset of prompt radio emission in some TDEs may be too early to be plausibly explained by accretion onto the SMBH, suggesting that other phases of the disruption process may also be capable of powering fast outflows.

\item \emph{Our results strongly motivate systematic, consistent radio, optical, and X-ray monitoring campaigns of TDEs, beginning as soon as possible after discovery and extending to years post-disruption.} Rapid public classification of new TDEs is also essential to ensure adequate multi-wavelength coverage at both early and late times, which is the only way to properly characterize the rates and origins of prompt and delayed radio emission in TDEs. Such campaigns are underway and will be reported in future work.
\end{itemize}

As larger samples of TDEs are discovered and the time cost of intensive targeted follow up becomes prohibitive, all-sky radio surveys will play an increasingly important role in systematically monitoring the full TDE population. Our independent VLASS discovery of the dramatic rebrightening of ASASSN-15oi \citep{ham+24} demonstrates the value of unbiased radio surveys for uncovering unexpected behavior --- it is unlikely that late-rising radio emission in TDEs would have been explored at all, without the initial VLASS detection of this TDE. Our few VLASS detections (Appendix \ref{sec:radio}) confirm that $2-6$ years post-disruption may be a particularly fruitful time to search for late-rising low-frequency radio emission in optically selected TDEs.

All-sky radio surveys are now also uncovering new nuclear transients that may be TDEs missed at other wavelengths (e.g., \citealt{amh+20, rdc+22, srd+23,srd+23b, ddg+24}). Our results suggest that deep late-time X-ray observations of these ambiguous radio transients may be one way to gain insight even when optical and X-ray observations extending to several years prior to the radio discovery are not available. While VLASS is the deepest radio survey of the northern sky conducted to date, many of the TDE outflows observed in the targeted radio observations of our sample are too faint to be detected in VLASS for all but the nearest TDEs, suggesting that VLASS's limited sensitivity, frequency coverage, and time sampling may still be inadequate to fully characterize this previously unappreciated late phase of TDEs' radio evolution. The planned ngVLA combined with wide-field radio surveys with the SKA will greatly increase our sensitivity to both low-luminosity radio outflows from TDEs and to distant jetted TDEs \citep{drf+15,vbm18}, providing critical support for the large numbers of higher-redshift TDEs expected from Rubin Observatory and other upcoming optical and multi-wavelength surveys.

\begin{acknowledgments}
We thank Sjoert van Velzen, Adelle Goodwin, James Miller-Jones, and Nayana AJ for useful conversations. {We also thank the anonymous referee for helpful comments that have improved this work.}

The National Radio Astronomy Observatory (NRAO) is a facility of the U.S.
National Science Foundation operated under cooperative agreement by Associated Universities, Inc. This research has made use of the CIRADA cutout service at URL cutouts.cirada.ca, operated by the Canadian Initiative for Radio Astronomy Data Analysis (CIRADA). CIRADA is funded by a grant from the Canada Foundation for Innovation 2017 Innovation Fund (Project 35999), as well as by the Provinces of Ontario, British Columbia, Alberta, Manitoba and Quebec, in collaboration with the National Research Council of Canada, the US National Radio Astronomy Observatory and Australia’s Commonwealth Scientific and Industrial Research Organisation. In addition to Swift observations proposed by us, we have also used the Swift archive at \url{https://swift.gsfc.nasa.gov/archive/}.  The MeerKAT telescope is operated by the South African Radio Astronomy Observatory, which is a facility of the National Research Foundation, an agency of the Department of Science and Innovation.

KDA and CTC acknowledge support provided by the NSF through award
SOSPA9-007 from the NRAO and award AST-2307668. KDA gratefully acknowledges support from the Alfred P. Sloan Foundation.
NF acknowledges support from the National Science Foundation Graduate Research Fellowship Program under Grant No. DGE-2137419. 
BDM was supported in part by the National Science Foundation (grant No. AST-2009255). The Flatiron Institute is supported by the Simons Foundation. 
MN is supported by the European Research Council (ERC) under the European Union’s Horizon 2020 research and innovation programme (grant agreement No.~948381) and by UK Space Agency Grant No.~ST/Y000692/1. 
ER-R was supported in part by  the Heising-Simons Foundation, NSF (AST-2150255 and AST-2307710), Swift (80NSSC21K1409,80NSSC19K1391) and Chandra (22-0142). 
FDC acknowledges support from
the UNAM-PAPIIT grant IN113424.

This research was supported in part by grant no. NSF PHY-2309135 to the Kavli Institute for Theoretical Physics (KITP). RC and RM acknowledge the Aspen Center for Physics, which is supported by National Science Foundation grant PHY-2210452, for its hospitality while this work was completed.

\end{acknowledgments}

%

\vspace{5mm}
\facilities{VLA, MeerKAT, ATCA, Swift(XRT and UVOT), Chandra, XMM-Newton}


\software{astropy \citep{2013A&A...558A..33A,2018AJ....156..123A}, mosfit \citep{gnv+18}, Source Extractor \citep{1996A&AS..117..393B}}, CASA \citep{casa}, pwkit \citep{pwkit}



\appendix

\section{VLASS search for late-rising radio emission from known TDEs}\label{sec:radio}

The advent of the VLASS, the most sensitive radio survey of the full northern sky to date, provides an exciting new opportunity to search for late-rising radio emission from known TDE candidates. 
VLASS began on 2017 September 7 and is using the VLA to map 80\% of the radio sky in three epochs spanning seven years \citep{vlass}. All VLASS observations are taken at $2-4$ GHz (S-band) and the typical rms sensitivity of the single-epoch images is $\sim100$ $\mu$Jy. 

Given the newly-recognized prevalence of long-lasting radio emission in TDEs, all-sky radio surveys also open up a complementary method of \emph{discovering} new TDEs in the nearby Universe. Blind radio searches with VLASS and radio surveys with the SKA precursors are now finding dozens of nuclear transients that may be TDEs, but interpretation is difficult without multi-frequency radio follow up (e.g., \citealt{amh+20, rdc+22, srd+23,srd+23b, ddg+24}). The rates and higher average radio luminosities of transients discovered in these blind searches suggest that not all of them are drawn from the same underlying population as the TDEs studied in this work, but it is not clear if this reflects a true dichotomy or a bias in host environment -- for example, TDEs in dense gas and dust-rich environments may produce stronger radio emission \citep{gmm+17}, but may have weaker observed optical and X-ray emission due to extinction. VLASS allows us to easily compare the radio properties of optical TDEs to those of other nuclear transients.

\begin{figure}
\centering
    \includegraphics[width=.7\textwidth]{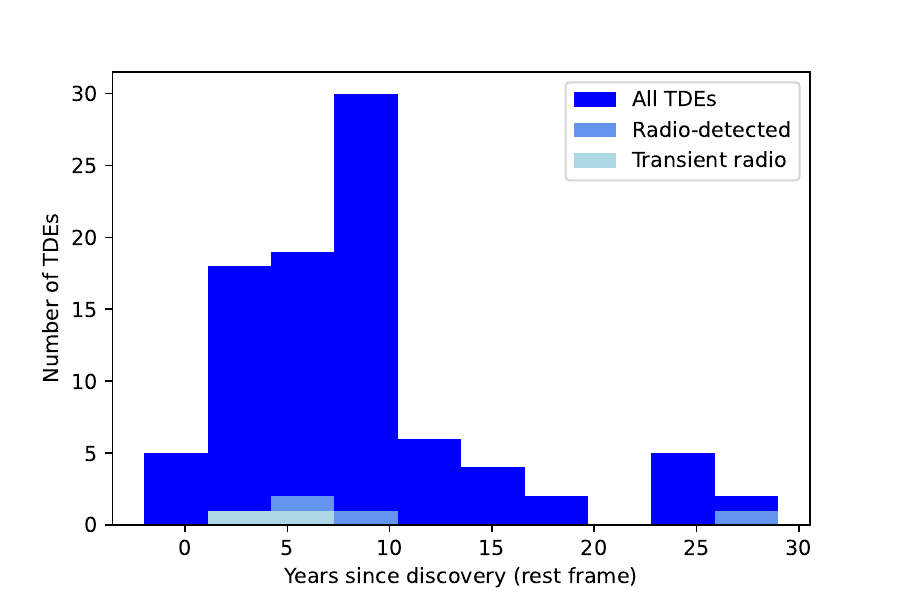}
    \caption{\label{fig:vlass}The ages of the TDEs in our VLASS sample at the time each was observed in VLASS epoch 1. We detect radio emission from 5 TDEs out of a total sample of \vlassnum. Three detections are likely host galaxy emission (NGC 5905, F01004-2237, and J100933), while two are confirmed to be transient radio emission associated with the TDE (Sw J1644+57 and ASASSN-15oi). Both confirmed TDE radio detections occur on timescales of $2-6$ years post-discovery (rest frame), suggesting that this may be an interesting timescale for deeper follow up. (A sixth TDE, AT2018hyz, is detected in VLASS epoch 3 at a similar timescale post-disruption.)} 
\end{figure}

Here, we use VLASS to investigate if the late-time radio properties of the optical TDEs studied in this work are representative of the broader, heterogeneously-selected full known TDE population, which includes TDEs discovered via a wide range of methods (e.g., as flares in all-sky X-ray or infrared surveys). We compiled a broad list of TDE candidates from the literature, consisting of objects from the Open TDE Catalog\footnote{The Open TDE Catalog, \url{https://github.com/astrocatalogs/tidaldisruptions}, provides a largely-complete list of TDEs reported in the literature through mid-2018. It includes some objects with ambiguous classification as well as transients widely accepted by the community as TDEs.} \citep{osc} and the TDE sample from \cite{fwl+20} that are within the VLASS footprint. Our target list is optimized for completeness rather than purity; we do not make any cuts based on TDE redshift, age, or discovery method. Our final sample contains \vlassnum\ objects discovered between 1990 and 2019, including 40 optical TDE candidates, 39 X-ray TDE candidates, and 18 TDE candidates identified via other methods, including nuclear infrared flares and bright transient gamma-ray emission (Figure \ref{fig:vlass}). The redshifts of our TDE candidates range from 0.000271 to 1.1853 (not including one X-ray transient in the Chandra Deep Field South with a photometric redshift $z\sim2$; \citealt{bts+17}); the median redshift of the sample is $z=0.088$. Approximately two thirds of the TDE candidates in this list had not received any published targeted radio follow up at the time of our search.

We cross-matched the locations of these \vlassnum\ TDE candidates with our Source Extractor catalog of possible VLASS source detections, publicly available {on Zenodo \citep{10.5281/zenodo.4895113}}. The creation of this catalog is described in \cite{stc+21}.  For this search, we utilized the Epoch 1 quick-look images provided by NRAO.\footnote{For additional information about the VLASS quick-look images and their limitations, see \url{https://science.nrao.edu/vlass/data-access/vlass-epoch-1-quick-look-users-guide}.} The first half of the data (VLASS1.1) was collected between September 2017 and February 2018 and suffers from additional uncertainties in the absolute flux calibration scale; the peak flux densities of sources fainter than 1 Jy are systematically low by 15\%, with an additional systematic scatter of $\sim\pm8$\%. The second half of the data (VLASS1.2) was collected between February and July 2019; the total flux densities of faint sources in these data are low by $\sim8$\% with a scatter of $\sim5\%$. As our primary goal is simply to determine the possible presence of transient radio emission, these calibration uncertainties are acceptable for our purposes.

We identified 5 TDEs located within 5 arcsec of a VLASS source detected with a signal-to-noise ratio $>5$ in our catalog, which are presented in Table \ref{tab:vlass}. We use the {\tt imfit} command within CASA to fit an elliptical Gaussian to each TDE and take the peak brightness to be equal to the flux density, as expected for an unresolved point source. Interestingly, each of these TDEs was discovered via a different method: NGC 5905 was one of the first TDE candidates ever reported and was identified as an X-ray flare from the center of a non-AGN galaxy \citep{bkd96}; Swift J1644+57 was discovered as a long-lasting, luminous gamma-ray and X-ray transient \citep{bgm+11,bkg+11,ltc+11}; ASASSN-15oi was discovered in an optical survey and classified as a TDE based on its nuclear location, photometric evolution, and spectroscopic properties \citep{hkp+16}; J100933 was discovered as a mid-infrared flare \citep{wyd+18}; and F01004-2237 was identified based on changes to its optical spectrum observed serendipitously during a survey of nearby ultra-luminous infrared galaxies \citep{tsr+17}. Two of these five objects, Swift J1644+57 and ASASSN-15oi, have well-studied radio counterparts previously reported in the literature \citep{zbs+11,bzp+12,wvl+12,zbm+13,ypv+16,ebz+18,cebp21,hca21,ham+24}. Swift J1644+57's radio emission, first detected within days of its discovery, is well-explained as synchrotron emission from an initially relativistic jet viewed on-axis (or slightly off-axis; see \citealt{bpm23}). Its detection in VLASS epochs 1 and 2 was previously reported by \cite{cebp21}. The epoch 1 VLASS detection of ASASSN-15oi, also reported by \cite{hca21} and \cite{ham+24}, was the first evidence for a dramatic second brightening of its radio emission, providing the initial impetus for the deeper late-time observations of other TDEs presented in \yvette\ and this work. NGC 5905 was the first TDE with a follow-up radio observation to search for jet emission. A VLA observation $\sim$ 6 years after the X-ray peak provided
an upper limit of $<0.15$ mJy at 8.46 GHz for a central point source \citep{kd01}. 
It was also observed as part of the NRAO VLA Sky Survey (NVSS). The NVSS detection of extended radio emission of NGC 5905 (21 mJy at 1.4 GHz) was found to be consistent with the starburst properties of the host galaxy
and consistent with pre-outburst extended radio emission \citep{k02,rdh+15}. 
The VLASS is less sensitive to extended emission due to the survey's higher resolution, but the VLASS flux densities of NGC 5905 are approximately constant in time and are also likely consistent with unrelated emission from the host galaxy. We include NGC 5905's VLASS flux density in Table \ref{tab:vlass} for completeness but do not investigate this source further. We also choose not to further observe F01004-2237, as the nature of this source is disputed. It was initially proposed to be a TDE \citep{tsr+17}, later re-classified as an unusual AGN Bowen fluorescence flare \citep{tar+19}, and finally recently suggested to be a repeating partial TDE (or two independent TDEs) after a second optical flare was observed in September 2021 \citep{sjd+24}. The continued activity of this source motivates future multi-wavelength monitoring to distinguish between the AGN and repeated or partial TDE scenarios.

We triggered additional multi-frequency follow up observations of the two most promising radio TDE candidates in our sample, ASASSN-15oi and J100933. We observed these targets with the VLA on 2020 May 20 (program 20A-492, PI: Alexander) and 2021 June 10 (program 21A-303, PI: Hajela). Observations were conducted at $1-12$ GHz (L, S, C, and X bands) when the VLA was in its C configuration. The data were reduced in CASA using standard procedures and flux densities were measured using the {\tt imtool fitsrc} command within the {\tt pwkit} package \citep{pwkit}. We have verified that this gives similar results to measuring the flux densities with {\tt imfit}. The follow up observations of ASASSN-15oi are presented in \cite{ham+24}. We include the measured flux densities of J100933 in Table \ref{tab:vlass} together with their statistical errors. The systematic uncertainty on the flux calibration for these observations is 5\% at all frequencies. In our 2021 observations, J100933 exhibits a rising spectrum with a peak around 9 GHz. 
We find that the flux density of J100933 at 3 GHz increased slightly between the time of the VLASS epoch 1 detection and our follow up observations two years later and then remained approximately constant for the next three years. Given the additional flux calibration uncertainties in the quick look images for epoch 1 of VLASS, this is roughly consistent with no significant temporal evolution over the full 5-year period. We thus conclude that the radio emission in this target is likely unrelated to the transient, although longer-term multi-frequency monitoring may be needed to confirm this due to the age of the TDE ($>11$ yr).   

We also visually inspected the coordinates of each of the \vlassnum\ TDEs in our sample in VLASS epochs 2.1, 2.2, and 3.1 quicklook data using the CIRADA cutout server to see if any of them appeared in the radio after VLASS epoch 1\@. We find only one new TDE undetected in VLASS epoch 1 that ``turns on'' in subsequent data: AT2018hyz, which is not detected in epochs 1 or 2 but appears as a bright point source in VLASS epoch 3. The VLASS data for AT2018hyz have already been presented in \cite{cba+22} and \yvette; we therefore do not repeat them here. We also continue to detect the five sources visible in our Epoch 1 search. NGC 5905, F01004-2237, and J100933 exhibit roughly constant radio brightness across all VLASS epochs, while Sw J1644+57 and ASASSN-15oi slowly fade. We include the flux densities of these five TDEs in all available VLASS epochs in Table \ref{tab:vlass}. 

In summary, our VLASS search of \vlassnum\ TDE candidates resulted in three detections of transient radio emission associated with genuine TDEs (two visible in all three VLASS epochs, and one that is only detected in VLASS epoch 3), two radio detections of TDE host galaxies, and one radio detection of a likely non-TDE transient (F01004-2237) or its host galaxy. Although our statistical constraining power is limited due to the small number of detections, the diverse multi-wavelength properties of these objects suggest no clear preference for optically selected TDEs among the TDE sub-population that is radio-bright at late times, at least at the bright end of the radio luminosity function. The low detection fraction seen in our search is consistent with early searches for late-time radio emission in TDEs, which concluded that powerful relativistic jets like that seen in the TDE Swift J1644+57 are rare \citep{k02,vfkf13}. Due to the lower sensitivity of available radio telescopes at the time and the relatively large distances of many early TDE candidates, these early searches were not sensitive to lower-luminosity radio emission similar to that seen in more recently-discovered nearby TDEs, many of which have peak radio luminosities $\nu L_{\nu} \lesssim 10^{38}$ erg s$^{-1}$ (\citealt{avhz20}; \yvette). Even our VLASS search can only detect $\nu L_{\nu} = 10^{38}$ erg s$^{-1}$ outflows out to $z\sim0.065$, meaning that such emission would be missed in $\gtrsim60$\% of our sample even if all TDEs were captured at or near their peak radio brightness (which is unlikely due to the broad distribution of TDE ages in our sample). Our results suggest that $\sim2-6$ years post-discovery may be a particularly fruitful timescale to search for late-rising radio emission in TDEs at GHz frequencies and demonstrates the value of long-duration monitoring, which will be enabled by future generations of more sensitive radio surveys.

\begin{deluxetable*}{lllllccl}
\tablenum{5}\label{tab:vlass}
\tablecaption{Radio observations of TDE candidates first detected in Epoch 1 of VLASS. We show both statistical errors from our source fitting procedure and systematic errors from the absolute flux density calibration scale of the data. Note that the VLASS errors are asymmetric to reflect that the peak flux densities of the quick look images are on average biased low relative to pointed observations (see text for details).}
\tablewidth{0pt}
\tablehead{
\colhead{TDE Name} & \colhead{Redshift} & \colhead{Earliest Detection} & \colhead{Observation} & \colhead{Program} & \colhead{Frequency} & \colhead{Flux density} & \colhead{Notes} \\
\nocolhead{Name} & \nocolhead{Number} & \colhead{(any wavelength)} & \colhead{Date} & \nocolhead{Program} & \colhead{(GHz)} & \colhead{(mJy $\pm$ stat $\pm$ sys)} & \nocolhead{(MJD)} } 

\startdata
NGC 5905* & 0.01124 & 1990 Jul 11 & 2017 Nov 11 & VLASS 1.1 & 3.0 & $1.59\pm0.12^{+0.08}_{-0.37}$ & X-ray TDE \\
 &  &  & 2020 Aug 3 & VLASS 2.1 & 3.0 & $1.96\pm0.16^{+0.10}_{-0.45}$ &  \\
 &  &  & 2023 Feb 9 & VLASS 3.1 & 3.0 & $1.65\pm0.12^{+0.08}_{-0.38}$ &  \\
\hline
F01004-2237* & 0.117835 & 2010 Jun 25 & 2018 Feb 13 & VLASS 1.1 & 3.0 & $3.47\pm0.13^{+0.17}_{-0.80}$ & AGN flare \\
 &  &  & 2020 Nov 1 & VLASS 2.1 & 3.0 & $3.39\pm0.15^{+0.17}_{-0.78}$ & \\ 
 &  &  & 2023 Jun 12 & VLASS 3.1 & 3.0 & $4.1\pm0.2^{+0.2}_{-0.9}$ & \\
\hline
Swift J1644+57 & 0.3543 & 2011 Mar 25 & 2017 Oct 7 & VLASS 1.1 & 3.0 & $1.32\pm0.11^{+0.07\dagger}_{-0.30}$ & gamma-ray TDE \\
 &  &  & 2020 Aug 4 & VLASS 2.1 & 3.0 & $0.64\pm0.16_{-0.15}^{+0.03\dagger}$ &  \\
 &  &  & 2023 Feb 11 & VLASS 3.1 & 3.0 & $0.346\pm0.08^{+0.02}_{-0.08}$ &  \\
\hline
ASASSN-15oi & 0.02 & 2015 Aug 14 & 2019 Jul 1 & VLASS 1.2 & 3.0 & $9.0\pm0.2_{-1.0}^{+0.6\ddagger}$ & optical TDE \\
 &  &  & 2022 Feb 15 & VLASS 2.2 & 3.0 & $4.95\pm0.14^{+0.35}_{-0.54}$ & \\
\hline
J100933* & 0.0719 & 2009 Feb 20 & 2019 Apr 18 & VLASS 1.2 & 3.0 & $1.88\pm0.15^{+0.13}_{-0.21}$ & mid-IR flare  \\
 &  &  & 2021 Jun 10 & 21A-303 & 1.263	& $1.59\pm0.16\pm0.08$ &   \\
 &  &  & 2021 Jun 10 & 21A-303 & 1.775	& $2.15\pm0.10\pm0.11$ &   \\
 &  &  & 2021 Jun 10 & 21A-303 & 2.5	& $2.49\pm0.04\pm0.12$ &   \\
 &  &  & 2021 Jun 10 & 21A-303 & 3.5	& $3.04\pm0.03\pm0.15$ &   \\
 &  &  & 2021 Jun 10 & 21A-303 & 3.0 & $2.85\pm0.04\pm0.14$ &  full S band image to match    \\
 &  &  &  &  & &  & VLASS (combines above 2 rows)  \\
 &  &  & 2021 Jun 10 & 21A-303 & 5.0	& $3.65\pm0.03\pm0.18$ &   \\
 &  &  & 2021 Jun 10 & 21A-303 & 7.0	& $4.12\pm0.04\pm0.21$ &   \\
 &  &  & 2021 Jun 10 & 21A-303 & 9.0	& $4.20\pm0.03\pm0.21$ &   \\
 &  &  & 2021 Jun 10 & 21A-303 & 11.0	& $4.06\pm0.04\pm0.20$ & \\
 &  &  & 2021 Dec 1 & VLASS 2.2 & 3.0 & $2.47\pm0.11^{+0.17}_{-0.27}$ & \\
 &  &  & 2024 Apr 30 & VLASS 3.2 & 3.0 & $2.46\pm0.13^{+0.17}_{-0.27}$ & \\
\enddata
\smallskip
*indicates origin of radio emission is either ambiguous or unrelated to the TDE (see text).

$^{\dagger}$ Consistent with values previously reported in \cite{cebp21}.

$^{\ddagger}$ Consistent with value previously reported in \cite{ham+24}.
\end{deluxetable*}

\section{Data Tables}\label{sec:data}

In this section, we present a table listing the original references for all published optical/UV, X-ray, and radio data used in our analysis (Table \ref{tab:data}) and complete data tables for all new observations. Our new radio observations of the TDE AT2019qiz (Section \ref{sec:rdata}) are listed in Table \ref{tab:19qiz} and new Swift-XRT observations of our sample (Section \ref{sec:xrays}) are listed in Table \ref{tab:xrays}. Table \ref{tab:xrays} also includes public archival Swift observations of our TDEs that have not previously been published, which were reduced using the same method as our new observations (Section \ref{sec:xrays}). We will also make the full multi-wavelength dataset for each TDE in our sample publicly available online through OTTER {(the Open mulTiwavelength Transient Event Repository; \citealt{fag+25}). OTTER has both a web application and a Python API for easy data access}.

{The} radio data first published in \yvette\ were taken with the VLA (programs 20B-492, PI: Alexander; 21-303, PI: Hajela; 22B-205, PI: Cendes), MeerKAT (programs SCI-20210212-YC-01, DDT-20220414-YC-01, and SCI-20220822-YC-01; PI: Cendes), and the Australia Telescope Compact Array (ATCA; programs C3472, PI: Cendes and C3325, PI: Alexander). The full list of observations, measured radio flux densities and upper limits, and details of the data reduction for these 24 objects are given in \cite{cba+22}, \yvette, and references therein. Our new radio observations of AT2019qiz are described in Section \ref{sec:rdata}; our measured flux densities and the observing program used to collect each data point are listed in Table \ref{tab:19qiz}. 

Sixteen of our TDEs were included in the uniform X-ray analysis published by \cite{ggy+23}, which includes observations taken by \textit{XMM-Newton} and by the X-ray Telescope (XRT, \citealt{Burrows05}) onboard the Neil Gehrels Swift Observatory \citep{Gehrels04}. We use the data from \cite{ggy+23} for 14 of these objects; for ASASSN-15oi, we instead use the more comprehensive dataset presented by \cite{ham+24}, and for AT2019qiz, we use the X-ray dataset from \cite{npm+24}, which corrects the XRT photometry for the presence of a nearby background source unrelated to the TDE. We also include published observations from the Chandra X-ray Observatory for a few TDEs: ASASSN-14ae \citep{jsg+20}, PS16dtm \citep{pli+23}, AT2017eqx \citep{nbb+19}, and AT2018hyz \citep{cba+22}. For AT2018fyk, we also include the late-time X-ray dataset from \cite{pcg+24} consisting of XRT, XMM, and Chandra observations at $t>1000$ d post-discovery. For ASASSN-19bt, we report \textit{XMM-Newton} photometry from \cite{hva+19} and Swift-XRT and Chandra photometry from \cite{caw+24}. 

We include all of the optical/UV photometry from \cite{nlr+22} for the TDEs presented in that work (Table \ref{tab:data} also lists the original references for these data). For AT2018lna, AT2019teq, AT2020mot, AT2020opy, AT2020pj, and AT2020wey we include $g$ and $r$-band photometry from the Zwicky Transient Facility (ZTF; \citealt{Bellm19}) obtained from the Automatic Learning for the Rapid Classification of Events (ALeRCE) broker \citep{Forster20} and UVOT photometry from \cite{hvg+23}.  For AT2020neh, AT2020nov, and AT2020vwl, we include only $g$ and $r$-band photometry from ZTF from ALeRCE. For OGLE17aaj we include I-band photometry from Optical Gravitational Lensing Experiment (OGLE; \citealt{wkk+14}) up until MJD = 58200. We include the UVOT photometry from \cite{Wevers19} for AT2018fyk and from \cite{vgh+21} for AT2018bsi. For ASASSN-19bt, we include both published photometry and the public light curve from the ASAS-SN Sky Patrol \citep{hsh+23}. Six TDEs in our sample (ASASSN-14li, ASASSN-15oi, AT2018hyz, ASASSN-19bt, AT2019teq, and AT2019ehz) show late-time plateaus in their optical and UV light curves that are not well-captured by MOSFiT, as discussed in Section \ref{sec:mosfit}. We therefore only use the early photometry from these events for our modeling. The cutoff date for each of these TDEs is listed in Table \ref{tab:data}. Similarly, for PS16dtm we include only the photometry from \cite{Blanchard17} as the additional photometry published by \cite{pli+23} is not well-captured by MOSFiT.

\startlongtable
\begin{deluxetable}{llll} 
\tablenum{6}
\tablecaption{References for all data used in our analysis. All data for which the reference is ``This work'' are given in Tables \ref{tab:19qiz} and \ref{tab:xrays}.\label{tab:data}}
\tablewidth{0pt}
\tablehead{
\colhead{TDE Name} & \colhead{References for Optical/} & \colhead{References for} & \colhead{References for} \\
\nocolhead{Name} & \colhead{UV photometry; time range} & \colhead{X-ray data} & \colhead{radio data} 
}
\startdata
ASASSN-14ae  & \cite{hpb+14}; & \cite{jsg+20}; & \yvette \\
  & \cite{hhs+21}; & This work & \\ 
  & \cite{nlr+22} &  &  \\
\hline
DES14C1kia & None & This work & \yvette \\
\hline
ASASSN-14li  & \cite{jgl+14}; & \cite{ggy+23}; & \cite{abg+16} \\
  & \cite{hkp+16a}; & This work &  \\
  & \cite{hhs+21}; & & \\
  & \cite{nlr+22}; &  &  \\
  & $t\leq57300$ MJD &  &  \\
\hline
iPTF15af  & \cite{bck+19}; & This work & \cite{bck+19}; \\
  & \cite{nlr+22} &  & \yvette \\
\hline
ASASSN-15oi & \cite{hkp+16};  & \cite{ham+24} & \cite{hca21}; \\
 & \cite{hhs+21}; &  & \cite{ham+24} \\
 & \cite{nlr+22};  &  &  \\
 & $t\leq57400$ MJD &  &  \\
\hline
iPTF16axa  & \cite{hgb+17}; &  This work & \yvette \\
  & \cite{hhs+21}; &   &  \\
  & \cite{nlr+22} &   &  \\
\hline
PS16dtm  & \cite{Blanchard17} & \cite{pli+23};  & \cite{Blanchard17}; \\
 & & This work & \yvette \\
\hline
iPTF16fnl  & \cite{bgh+17}; &  This work & \cite{bgh+17};  \\
  & \cite{hhs+21}; &   & \cite{hsf+21}; \\
  & \cite{nlr+22} &   &  \yvette \\
\hline
OGLE17aaj & \cite{wkk+14}; & This work & \cite{slb17}; \\
 & \cite{ghw+19}; &  &  \yvette \\
 & $t\leq58200$ MJD &  &  \\
\hline
AT2017eqx & \cite{nbb+19}; & \cite{nbb+19} (Chandra);  & \cite{nbb+19};  \\
 & \cite{hhs+21}; & This work (XRT) &  \yvette \\
 & \cite{nlr+22} &  & \\
\hline
AT2018zr & \cite{vgc+19}; & \cite{ggy+23} & \cite{vgc+19}; \\ 
 & \cite{hhs+19}; &  & \yvette \\
 & \cite{vgh+21}; &  &  \\
 & \cite{hhs+21}; & &  \\
 & \cite{nlr+22} & &  \\
\hline
AT2018bsi & \cite{vgh+21} & \cite{ggy+23}; & \yvette \\
& & This work & \\
\hline
AT2018dyb & \cite{lda+19}; & This work & \cite{hat+20}; \\
 & \cite{hat+20}; &  & \yvette \\
 & \cite{hhs+21}; &  &  \\
 & \cite{nlr+22} &  &  \\
\hline
AT2018fyk & \cite{Wevers19} & \cite{ggy+23}; & \cite{Wevers19};\\
 &  & \cite{pcg+24} & \cite{wpv+21};  \\
 &  &  & \yvette;  \\
 &  &  & \cite{cba+24b} \\
\hline
AT2018hco & \cite{vgh+21}; &  \cite{ggy+23};  & \yvette \\ 
& \cite{hhs+21}; &  This work & \\ 
& \cite{nlr+22} &   &  \\ 
\hline
AT2018hyz & \cite{gns+20}; &  \cite{cba+22}; & \cite{hsbf18};\\
 & \cite{vgh+21}; &  \cite{ggy+23} & \cite{gns+20};  \\
 & \cite{hhs+21}; &   & \cite{cba+22}; \\
 & \cite{nlr+22};  &   & \yvette \\
 & $t\leq58800$ MJD &   &  \\
\hline
AT2018lna & \cite{vgh+21}; & \cite{ggy+23}; & \yvette  \\
& \cite{hhs+21}; & This work &  \\
& \cite{nlr+22}; &  &  \\
& \cite{hvg+23} &  &  \\
\hline
ASASSN-19bt & ASAS-SN Sky Patrol; & \cite{hva+19}; & \cite{caw+24} \\
 & \cite{hva+19}; & \cite{caw+24} &  \\
 & \cite{hhs+21};  &  &  \\
 & \cite{nlr+22};  &  &  \\
 & $t\leq58650$ MJD &  &  \\
\hline
AT2019azh & \cite{vgh+21}; & \cite{ggy+23}; & \cite{gvm+22}; \\ 
 & \cite{hhs+21}; & This work & \cite{shf+22} \\ 
 & \cite{nlr+22} &  &  \\ 
\hline
AT2019dsg & \cite{vgh+21}; & \cite{ggy+23};  & \cite{cab+21};  \\
 & \cite{cwj+21}; & This work & \cite{svk+21};  \\
 & \cite{hhs+21}; & & \yvette \\
 & \cite{nlr+22} &  &  \\
\hline
AT2019ehz & \cite{vgh+21}; & \cite{ggy+23} & \yvette \\
 & \cite{hhs+21}; &  &  \\
 & \cite{nlr+22}; &  & \\
 & $t\leq58800$ MJD &  & \\
\hline
AT2019eve & \cite{vgh+21} & This work & \yvette \\
 & \cite{hhs+21} &  & \\
 & \cite{nlr+22} &  & \\
\hline
AT2019qiz & \cite{vgh+21}; & \cite{npm+24} & This work;  \\
 & \cite{hhs+21}; &  & Franz et al. (in prep) \\
 & \cite{nwo+20}; &  &  \\
 & \cite{nlr+22} &  &  \\
\hline
AT2019teq & ZTF $g,r$ (ALERCE); & \cite{ggy+23}; & \cite{ATelCendes2022}; \\
 & \cite{hvg+23}; & This work & \yvette \\
 & $t\leq58850$ MJD & & \\
\hline
AT2020pj & ZTF $g,r$ (ALERCE); & \cite{ggy+23};  & \yvette \\
& \cite{hvg+23} & This work &  \\
\hline
AT2020mot & ZTF $g,r$ (ALERCE);  & This work & \cite{lkb+23}; \\
 & \cite{hvg+23} & & \yvette \\
\hline
AT2020neh & ZTF $g,r$ (ALERCE) & This work & \cite{abm+22}; \\
 &  &  & \yvette \\
\hline
AT2020nov & ZTF $g,r$ (ALERCE) & This work & \yvette \\
\hline
AT2020opy & ZTF $g,r$ (ALERCE); & This work & \cite{gmv+23} \\
 & \cite{hvg+23} &  & \\
\hline
AT2020wey & ZTF $g,r$ (ALERCE); & \cite{ggy+23}; & \yvette \\
 & \cite{hvg+23} & This work & \\
\hline
AT2020vwl & ZTF $g,r$ (ALERCE) & \cite{ggy+23}; & \cite{gam+23} \\
 &  & This work & \\
\enddata
\end{deluxetable}

\begin{deluxetable*}{lcccc}
\tablenum{7}
\tablecaption{New radio observations of AT2019qiz included in this work. We show both statistical-only errors from our fitting procedure and additional systematic errors from the VLA's absolute flux calibration. \label{tab:19qiz}}
\tablewidth{0pt}
\tablehead{
\colhead{MJD} & \colhead{Time since} & \colhead{Flux density} & \colhead{Frequency} & \colhead{Program} \\
\nocolhead{Name} &  \colhead{Discovery (d)} & \colhead{(mJy $\pm$ stat $\pm$ sys)} & \colhead{(GHz)} & \nocolhead{Program} 
}
\startdata
58752.41169	 & 	6.91	 & $	0.033	\pm	0.008	\pm	0.002	$ & 	15.0	 & 	VLA/19A-013	 \\
58773.36964	 & 	27.87	 & $	0.093	\pm	0.012	\pm	0.005	$ & 	9.0	 & 	VLA/19A-013	 \\
58799.36227	 & 	53.86	 & $	0.15	\pm	0.04	\pm	0.007	$ & 	5.0	 & 	VLA/19A-013	 \\
58799.36227	 & 	53.86	 & $	0.65	\pm	0.03	\pm	0.03	$ & 	9.0	 & 	VLA/19A-013	 \\
58823.29204	 & 	77.79	 & $	0.40	\pm	0.04	\pm	0.02	$ & 	5.0	 & 	VLA/19A-013	 \\
58823.29204	 & 	77.79	 & $	1.33	\pm	0.02	\pm	0.07	$ & 	9.0	 & 	VLA/19A-013	 \\
58871.11659	 & 	125.62	 & $	1.41	\pm	0.03	\pm	0.07	$ & 	5.0	 & 	VLA/19A-013	 \\
58871.11659	 & 	125.62	 & $	2.40	\pm	0.02	\pm	0.12	$ & 	9.0	 & 	VLA/19A-013	 \\
58981.73736	 & 	236.24	 & $	1.64	\pm	0.05	\pm	0.08	$ & 	5.0	 & 	VLA/20A-372	 \\
58981.73736	 & 	236.24	 & $	0.81	\pm	0.10	\pm	0.04	$ & 	9.0	 & 	VLA/20A-372	 \\
59213.07616	 & 	467.58	 & $	1.366	\pm	0.015	\pm	0.07	$ & 	5.0	 & 	VLA/20A-372	 \\
59213.07616	 & 	467.58	 & $	0.695	\pm	0.013	\pm	0.03	$ & 	9.0	 & 	VLA/20A-372	 \\
59378.77265	 & 	633.27	 & $	0.48	\pm	0.06	\pm	0.02	$ & 	5.0	 & 	VLA/21A-303	 \\
59378.77265	 & 	633.27	 & $	0.260	\pm	0.017	\pm	0.013	$ & 	9.0	 & 	VLA/21A-303	 \\
59916	 & 	1170.50	 & $	0.618	\pm	0.015	\pm	0.03	$ & 	1.285	 & 	MKT-22085	 \\
60028.00797	 & 	1282.51	 & $	0.7	\pm	0.2	\pm	0.02	$ & 	3.0	 & 	VLASS 3.1	  \\
\enddata
\end{deluxetable*}

\startlongtable
\begin{deluxetable*}{lcccc}
\tablenum{8}
\tablecaption{New Swift-XRT data used in this work. This table includes both our new late-time TOO observations and public archival observations re-reduced by us. The date column gives the mean observation time for each data point, while the uncertainty on the time since discovery is taken to be either the observation length for a single observation or the full time range for stacked observations. The luminosity is calculated assuming a $\Gamma=2$ power-law spectrum, as discussed in Section \ref{sec:xrays}. All upper limits are $3\sigma$ (Gaussian-equivalent). {This is a subset of the full table, which will be available online in machine-readable format.}\label{tab:xrays}}
\tablewidth{0pt}
\tablehead{
\colhead{TDE Name} & \colhead{Date} & \colhead{Time since} & \colhead{0.3-10 keV Count Rate} & \colhead{0.3-10 keV Luminosity} \\
\nocolhead{Name} & \colhead{(MJD)} & \colhead{discovery (d)} & \colhead{($10^{-3}$ counts s$^{-1}$)} & \colhead{($10^{41}$ erg s$^{-1}$)} 
}
\startdata
ASASSN14ae & 59590.754 & $2908.2^{+0.2}_{-8.1}$ & $1.3^{+0.6}_{-0.4}$ & $3.1^{+1.3}_{-1.0}$ \\
ASASSN14ae & 56779.066 & $97^{+660}_{-94}$ & $<0.401$ & $<0.955$ \\
ASASSN14ae & 59891.566 & $3209.06^{+0.17}_{-0.17}$ & $2.1^{+1.4}_{-1.0}$ & $5^{+3}_{-2}$ \\
ASASSN14ae & 59861.832 & $3179.32^{+0.02}_{-0.59}$ & $<5.95$ & $<14.2$ \\
ASASSN14ae & 60419.68 & $3737.168^{+0.003}_{-0.003}$ & $<16.5$ & $<39.3$ \\
ASASSN14ae & 60502.559 & $3820.0^{+0.2}_{-0.6}$ & $<10.3$ & $<24.6$ \\
ASASSN14ae & 60591.594 & $3909.1^{+0.4}_{-0.5}$ & $<4.51$ & $<10.7$ \\
\enddata
\end{deluxetable*}




\bibliography{sample631}{}
\bibliographystyle{aasjournal}



\end{document}